\documentclass[prd,twocolumn,nofootinbib,reprint,preprintnumbers,showkeys]{revtex4-1}
\usepackage{color}
\usepackage{amsmath,amsfonts,bm}
\usepackage{graphicx, epstopdf,scrextend}
\usepackage{mathrsfs,mathtools}

\usepackage{pgfplots,pgfplotstable}
\usepgfplotslibrary{fillbetween}
\usepgfplotslibrary{groupplots}
\usetikzlibrary{intersections}
\pgfplotsset{compat=newest}

\pgfplotsset{every axis/.style={
    width=8.6cm,
    height=8.6cm,
    grid=both,
    scaled ticks=false,
    yticklabel style={/pgf/number format/.cd, fixed,precision=5}
  }
}

\definecolor{darkgreen}{rgb}{0,0.4,0} 
\definecolor{darkblue}{rgb}{0,0,0.6} 
\usepackage[colorlinks=true,linkcolor=darkblue,citecolor=darkgreen]{hyperref}

\newcommand{\as}{\alpha_{\mathrm{s}}}

\newcommand{\LA}{\mathrm{A}}

\newcommand{\LF}{\mathrm{F}}

\newcommand{\LR}{\mathrm{R}}

\newcommand{\LT}{\mathrm{T}}

\newcommand{\La}{\mathrm{a}}
\newcommand{\Lb}{\mathrm{b}}
\newcommand{\Lc}{\mathrm{c}}

\newcommand{\Lg}{\mathrm{g}}

\newcommand{\Ls}{\mathrm{s}}

\newcommand{\LCP}{{\textsc{lc}\raisebox{+0.5 pt}{$\scriptscriptstyle{+}$}}}

\newcommand{\cH}{\mathcal{H}}

\newcommand{\cN}{\mathcal{N}}
\newcommand{\cO}{\mathcal{O}}

\newcommand{\cU}{\mathcal{U}}
\newcommand{\cV}{\mathcal{V}}

\newcommand{\cX}{\mathcal{X}}

\newcommand{\scE}{\textsc{e}}

\newcommand{\GeV}{\ \mathrm{GeV}}
\newcommand{\TeV}{\ \mathrm{TeV}}

\definecolor{red}{rgb}{1,0,0}
\newcommand{\NOTE}[1]{\textcolor{red}{ \bf[NOTE: #1]}}

\definecolor{darkgreen}{rgb}{0,0.4,0}

\def\mi{{\mathrm i}}

\def\ket#1{\big|{#1}\big\rangle}
\def\bra#1{\big\langle{#1}\big|}
\def\brax#1{\big\langle{#1}}   
\def\<>#1{\big\langle{#1}\big\rangle}
\def\[]#1{\big[{#1}\big]}

\def\sket#1{\big|{#1}\big)}
\def\sbra#1{\big({#1}\big|}
\def\sbrax#1{\big({#1}}        

\newbox\charbox
\newbox\slabox
\def\s#1{{      
        \setbox\charbox=\hbox{$#1$}
        \setbox\slabox=\hbox{$/$}
        \dimen\charbox=\ht\slabox
        \advance\dimen\charbox by -\dp\slabox
        \advance\dimen\charbox by -\ht\charbox
        \advance\dimen\charbox by \dp\charbox
        \divide\dimen\charbox by 2
        \raise-\dimen\charbox\hbox to \wd\charbox{\hss/\hss}
        \llap{$#1$}
}}

\newif\ifusefigs
\usefigstrue

\begin{document}

\title{Exponentiating virtual imaginary contributions in a parton shower}

\author{Zolt\'an Nagy}

\affiliation{
  DESY,
  Notkestrasse 85,
  22607 Hamburg, Germany
}

\email{Zoltan.Nagy@desy.de}

\author{Davison E. Soper}

\affiliation{
Institute of Theoretical Science,
University of Oregon,
Eugene, OR  97403-5203, USA
}

\email{soper@uoregon.edu}

\begin{abstract}
The operator in a parton shower algorithm that represents the imaginary part of virtual Feynman graphs has a non-trivial color structure and is large because it is proportional to a factor of $4\pi$. In order to improve the treatment of color in a parton shower, it may help to exponentiate this phase operator. We show that it is possible to do so by exponentiating matrices that are no larger than $14\times14$. Using the example of the probability to have a gap in the rapidity interval between two high transverse momentum jets, we test this exponentiation algorithm by comparing to the result of treating the phase operator perturbatively. We find that the exponentiation works, but that the net effect of the exponentiated phase operator is quite small for this problem, so that one can as well use the perturbative approach. 
\end{abstract}

\keywords{perturbative QCD, parton shower}
\preprint{DESY 19-143}
\date{29 August 2019}

\maketitle

\section{\label{sec:intro}Introduction}

There have been efforts to improve the accuracy of leading order parton shower programs by moving away from the leading color (LC) approximation in the treatment of QCD color \cite{NScolor, PlatzerSjodahl, NSColorEffects, Seymour2018, PlatzerSjodahlThoren, Isaacson:2018zdi, Forshaw:2019ver, NSColorTheory, NSGapColor}. For this effort to be consistent with quantum mechanics, one needs to work with bra and ket color amplitudes and to implement the color content of virtual graphs acting on these amplitudes \cite{NScolor, NSColorEffects, Seymour2018, Forshaw:2019ver, NSColorTheory, NSGapColor}.

In the paper \cite{NSColorTheory}, we explored how to include QCD color in a parton shower in a systematically improvable way. Then in Ref.~\cite{NSGapColor} we applied this algorithm, as implemented in our program \textsc{Deductor}, to the problem of gaps \cite{AtlasGap, HEJgap, HocheSchonherr, Manchester2009, EarlyGap1, EarlyGap2, Manchester2005, NonGlobal1, NonGlobal2, NonGlobal3, NonGlobal4, SuperLeading1, SuperLeading2, Manchester2011} in the rapidity interval between two high transverse momentum jets. In this paper, we explore the possibility of exponentiating the contributions from the imaginary part of virtual graphs, which are proportional to $\mi\pi$ times a non-trivial color operator.

The basic idea in Ref.~\cite{NSColorTheory} is simple. In an ideal, but not practical, leading order parton shower with exact color, the shower is represented as an operator $\cU(t,t_0)$. This operator takes the shower from an initial stage at a hard momentum scale, corresponding to a shower time $t_0$, to a later stage with a softer momentum scale, corresponding to a shower time $t$ \cite{NSI, NSAllOrder}. The operator $\cU(t,t_0)$ acts on a linear space of states $\sket{\rho}$, the {\em statistical space}. Ignoring spin but accounting for color, this space consists of functions of partonic momentum and flavor variables together with a density matrix in the space of the colors of the partons. The partonic momentum and flavor variables, $\{p,f\}_m$, consist of the number $m$ of final state partons and the momenta $p$ and flavors $f$ of the partons (including two initial state partons).  We can expand $\rho(\{p,f\}_m)$ in color basis states,
\begin{equation}
\label{eq:rhomatrix}
\rho(\{p,f\}_m) = \sum_{\{c,c'\}_m}
\rho(\{p,f,c,c'\}_m)
\ket{\{c\}_m}\bra{\{c'\}_m}
\;.
\end{equation}
Thus we work with amplitudes in parton color space, with a ket amplitude $\ket{\{c\}_m}$ and a conjugate bra amplitude $\bra{\{c'\}_m}$. To get the probability that corresponds to $\sket{\rho}$, we form $\sbrax{1}\sket{\rho}$, which is an instruction to sum over the number of partons, integrate over their momenta, sum over their flavors, and take the color trace of $\rho(\{p,f\}_m)$, obtaining inner products $\brax{\{c'\}_m}\ket{\{c\}_m}$.

With exact color, the operator $\cU(t,t_0)$ is the solution of the evolution equation
\begin{equation}
\label{eq:evolutionU}
\frac{d}{dt}\,\cU(t,t_0)
= [\cH_I(t) - \cV(t)]\,\cU(t,t_0)
\;.
\end{equation}
Here $\cH_I(t)$ is an operator that generates parton splittings (at first order in $\as$), increasing the number of partons by one each time it acts, and $\cV(t)$ is an operator that approximates virtual graphs and leaves the number of partons unchanged. The operators $\cH_I(t)$ and $\cV(t)$ have complicated color structures, so that it is not practical to implement Eq.~(\ref{eq:evolutionU}) directly. There is an approximation, the LC+ approximation, that generalizes the leading color approximation \cite{NScolor}. In the LC+ approximation, we replace $\cH_I(t)$ and $\cV(t)$ by approximate operators $\cH^\LCP(t)$ and $\cV^\LCP(t)$. Then we solve
\begin{equation}
\label{eq:evolutionLCplus}
\frac{d}{dt}\,\cU^\LCP(t,t_0)
= [\cH^\LCP(t) - \cV^\LCP(t)]\,
\cU^\LCP(t,t_0)
\;.
\end{equation}
The solution takes the form
\begin{equation}
  \label{eq:evolutionsolutionLCplus}
  \begin{split}
    \cU^\LCP{}&(t,t_0) =
    \cN^\LCP(t,t_0)
    \\
    &+
    \int_{t_0}^t\!d\tau\ 
    \cU^\LCP(t,\tau)\,
    \cH^\LCP(\tau)\,
    \cN^\LCP(\tau,t_0) 
    \;.
  \end{split}
\end{equation}
Here $\cN^\LCP(t_{2},t_1)$ is the no-splitting operator,
\begin{equation}
\label{eq:NLCplus}
\cN^\LCP(t_2,t_1) = \exp\left[
-\int_{t_1}^{t_2} d\tau\ \cV^\LCP(\tau)\right]
\;.
\end{equation}
The operator $\cV^\LCP(\tau)$ is diagonal in the color basis that we use, the trace basis, so that it is practical to calculate its exponential. Then the diagonal matrix elements of $\cN^\LCP(t_2,t_1)$ constitute the Sudakov factor that is needed to choose the next splitting time in the parton shower algorithm.

Now, the LC+ approximation leaves out operators $\Delta \cH(t)$ and $\Delta \cV(t)$,
\begin{equation}
\begin{split}
\label{eq:DeltaHDeltaV}
\Delta \cH(t) ={}& \cH_I(t) - \cH^\LCP(t)
\;,
\\
\Delta \cV(t) ={}& \cV(t) - \cV^\LCP(t)
\;.
\end{split}
\end{equation}
The operator $\Delta\cV(t)$ contains two terms:
\begin{equation}
\label{eq:DeltaVreDeltaVipi}
\Delta \cV(t) = \Delta \cV_\textrm{Re}(t) + \cV_{\mi\pi}(t)
\;.
\end{equation}
The first term, $\Delta \cV_\textrm{Re}(t)$, approximates the terms in the real part of virtual graphs that are not included in $\cV^\LCP(t)$. The second term, $\cV_{\mi\pi}(t)$, corresponds to the imaginary part of virtual graphs.  

In order to set up the general expansion in powers of $\Delta \cH(t)$, $\Delta \cV_\textrm{Re}(t)$, and $\cV_{\mi\pi}(t)$, we first define an operator $\cX(t,t_0)$ by
\begin{equation}
  \label{eq:evolutionX}
  \begin{split}
    \cX&(t,t_0) =
    \\
    &
     1- \int_{t_0}^t\!d\tau\ 
    \cX(t,\tau)\,
    \cN^{\LCP}(t,\tau)\,
    \Delta \cV(\tau)\,
    \cN^\LCP(t,\tau)^{-1}
    \;.
  \end{split}
\end{equation}
This equation is to be solved iteratively to give a series in powers of $\Delta \cV$:
\begin{equation}
  \begin{split}
    \label{eq:NDeltaexpansion}
    \cX{}&(t,t_0) = 
    \\ & 
    1
    - \int_{t_0}^t\!d\tau\ 
    \cN^\LCP(t,\tau)\,
    \Delta \cV(\tau)\,
    \cN^\LCP(t,\tau)^{-1}
    \\ & 
    + \cdots 
    \;.
  \end{split}
\end{equation}
Then the evolution equation for $\cU(t,t_0)$ becomes
\begin{equation}
  \label{eq:evolutionU5}
  \begin{split}
    \cU{}&(t,t_0) = 
    \cX(t,t_0)\, \cN^\LCP(t,t_0)
    \\&
    + \int_{t_0}^t\!d\tau\ 
    \cU(t,\tau)\,
    \big[\cH^\LCP(\tau) + \Delta \cH(\tau)\big]\,
    \\
    &\qquad\times\cX(\tau,t_0)\, \cN^\LCP(\tau,t_0) 
    \;.
  \end{split}
\end{equation}
This is also to be solved iteratively. 

The iterative expansions of Eqs.~(\ref{eq:NDeltaexpansion}) and (\ref{eq:evolutionU5}) give $\cU(t,t_0)$ as series in powers of $\Delta \cH(t)$, $\Delta \cV_\textrm{Re}(t)$, and $\cV_{\mi\pi}(t)$. Then we can specify non-negative integers $N_\textrm{Re}$ and $N_{\mi\pi}$ and retain contributions proportional to $[\Delta \cH]^A  [\Delta \cV_\textrm{Re}]^B [\cV_{\mi\pi}]^C$ with $A + B \le N_\textrm{Re}$, $C \le N_{\mi\pi}$ and $A + B + C \le \max\{N_\textrm{Re}, N_{\mi\pi}\}$.

In the present paper, we ask whether one could exponentiate $\cV_{\mi\pi}$ in Eq.~(\ref{eq:DeltaVreDeltaVipi}), so that we effectively set $N_{\mi\pi} = \infty$. If one could do this, one would have a parton shower that is well adapted for observables in which $\cV_{\mi\pi}$ is effectively a large operator. 

To include all powers of $\cV_{\mi\pi}$, we would like to replace $\cN^\LCP(t_2,t_1)$ in Eq.~(\ref{eq:NLCplus}) by
\begin{equation}
\label{eq:NscE}
\cN^{\scE}(t_2,t_1) = \mathbb{T}\exp\!\left[
-\int_{t_1}^{t_2}\! d\tau 
\left\{\cV^\LCP(\tau) + \cV_{\mi\pi}(\tau)\right\}\right]
.
\end{equation}
Here $\mathbb{T}$ indicates time ordering, with the latest shower times to the left. Then we would use not the full $\Delta \cV(t)$ in Eq.~(\ref{eq:evolutionX}), but only $\Delta \cV_\textrm{Re}(t)$:
\begin{align}
\label{eq:evolutionXnew}
\cX^\scE&(t,t_0) =
\\&
     1- \int_{t_0}^t\!d\tau\ 
    \cX^\scE(t,\tau)\,
    \cN^{\scE}(t,\tau)\,
    \Delta \cV_\textrm{Re}(\tau)\,
    \cN^\scE(t,\tau)^{-1}
\;.\notag
\end{align}
With $\cV_{\mi\pi}$ exponentiated, the evolution equation (\ref{eq:evolutionU5}) becomes
\begin{equation}
\label{eq:evolutionU6}
\begin{split}
    \cU{}&(t,t_0) = 
    \cX^\scE(t,t_0)\, \cN^\scE(t,t_0)
    \\&
    + \int_{t_0}^t\!d\tau\ 
    \cU(t,\tau)\,
    \big[\cH^\LCP(\tau) + \Delta \cH(\tau)\big]\,
    \\
    &\qquad\times\cX^\scE(\tau,t_0)\, \cN^\scE(\tau,t_0) 
    \;.
\end{split}
\end{equation}
%

\section{\label{sec:exponentiating}Exponentiating $\cV_{i\pi}$}

In order to examine whether it is practically possible to include $\cV_{\mi\pi}$ along with $\cV^\LCP$ in the no-splitting operator, as in Eq.~(\ref{eq:NscE}), we need to understand the color structure of $\cV_{\mi\pi}$. The structure of $\cV_{\mi\pi}$ (assuming massless partons) is given in Eq.~(10.14) of Ref.~\cite{NScolor},
\begin{equation}
\label{eq:Vipi}
\cV_{\mi\pi}(t) = - 4\mi \pi \frac{\as}{2\pi}
\big([(T_\La\cdot T_\Lb)\otimes 1]
- [1\otimes (T_\La\cdot T_\Lb)]\big)
\;.
\end{equation}
Here $T_\La$ represents the insertion of a color matrix $T^c$ on incoming parton line ``a'', $T_\Lb$ represents the insertion of a color matrix $T^c$ on incoming parton line ``b'', and the dot in $(T_\La\cdot T_\Lb)$ represents a summation over the octet color index $c$. In $[(T_\La\cdot T_\Lb)\otimes 1]$ the color matrices act on the ket state, while in $[1\otimes (T_\La\cdot T_\Lb)]$, they act on the bra state. When we take the color trace, we get $\sbra{1} \cV_{\mi\pi}(t) = 0$ because the two terms in Eq.~(\ref{eq:Vipi}) have opposite signs. 

At first sight, the goal of exponentiating $\cV_{\mi\pi}$ in a practical algorithm seems unreachable because the color operators in Eq.~(\ref{eq:Vipi}) act on a color space that has a very large dimensionality. In fact, an $n$ gluon statistical state vector is expanded in $[(n-1)!/2]^2$ basis vectors if we use the trace basis. For $n = 10$ this is $3\times 10^{10}$ basis vectors.

However, let us look at this more closely. When the operator $\cN^{\scE}(t_2,t_1)$ operates on a statistical space basis vector $\sket{\{p,f,c',c\}_m}$, it leaves the number of partons, their momenta, and their flavors unchanged. It acts as a matrix on the colors. However, this matrix is simpler than it would seem because it is a product of two matrices, one for the ket colors and one for the bra colors:
\begin{equation}
\begin{split}
\label{eq:Nmatrix}
\cN^{\scE}(t_2,t_1)&\sket{\{p,f,c',c\}_m} 
\\={}& 
\sum_{\{\hat c\}_m} N(t_2,t_1,\{p,f\}_m,\{\hat c\}_m,\{c\}_m )
\\
&\times \sum_{\{\hat c'\}_m} 
N^*(t_2,t_1,\{p,f\}_m,\{\hat c'\}_m,\{c'\}_m )
\\&\times
\sket{\{p,f,\hat c',\hat c\}_m}
\;.
\end{split}
\end{equation}
In order to simplify the notation, we can write these matrices as operators, an operator $n$ on the ket color space,
\begin{equation}
\begin{split}
\label{eq:Nton}
\sum_{\{\hat c\}_m} N(t_2,t_1, \{p,f\}_m,&\{\hat c\}_m,\{c\}_m )
\ket{\{\hat c\}_m}
\\
={}& n(t_2, t_1, \{p,f\}_m)\ket{\{c\}_m}
\end{split}
\end{equation}
and an operator $n^\dagger$ on the bra color space,
\begin{equation}
\begin{split}
\label{eq:Ntonstar}
\sum_{\{\hat c'\}_m} 
\bra{\{\hat c'\}_m}&
N^*(t_2,t_1, \{p,f\}_m,\{\hat c'\}_m,\{c'\}_m )
\\
={}& \bra{\{c'\}_m} n^\dagger(t_2,t_1, \{p,f\}_m)
\;.
\end{split}
\end{equation}

The operator $n$ has the form
\begin{equation}
\begin{split}
\label{eq:noperator}
n(t_2, t_1, &\{p,f\}_m)
\\ 
={}&
\mathbb{T}\exp\bigg[
-\int_{t_1}^{t_2} dt\ \frac{\as(t)}{2\pi}\,
a(t,\{p,f\}_m)
\bigg]
\;.
\end{split}
\end{equation}
Here $\mathbb{T}$ indicates ordering according to the shower time $t$. In $n(t_2,t_1,\{p,f\}_m)$, which operates on the ket color space, the latest shower times are on the left. In $n^\dagger(t_2,t_1, \{p,f\}_m)$, which operates on the bra color space, the latest shower times are on the right and the sign of the imaginary part of the exponent is reversed. 

The operator $a$ in the exponent of Eq.~(\ref{eq:noperator}) has the decomposition \cite{NSColorTheory}:
\begin{equation}
\begin{split}
\label{eq:adecomposition}
a(t,\{p,f\}_m) ={}& 
a^\LCP_\textrm{coll}(t,\{p,f\}_m)
+ a^\LCP_\textrm{soft}(t,\{p,f\}_m)
\\&-
4\mi\pi\, T_\La\cdot T_\Lb
\;.
\end{split}
\end{equation}

First, there is $a^\LCP_\textrm{coll}$. This operator comes from ``direct'' graphs, in which a parton is emitted from parton line $l$ and absorbed on the same parton line,
\begin{align}
\label{eq:acolldef}
\frac{\as(t)}{2\pi}\,&
a^\LCP_\textrm{coll}(t,\{p,f\}_m)
\\
&= 
\sum_{l}\int d\zeta\,
\delta\big(t - T_l(\zeta,\{p\}_{m})\big)
\lambda_{ll}(\{p,f\}_m, \zeta)\,
I_\Lc
\;.
\notag
\end{align}
There is a sum over the index of the emitting parton $l$. We integrate over momentum splitting variables $\Lambda$, $z$, $\phi$ and flavor splitting variables. Here $\Lambda$ is a hardness variable, $z$ is a momentum fraction, and $\phi$ is an azimuthal angle. The splitting variables are collectively denoted by $\zeta$. The function $T_l(\zeta,\{p\}_{m})$ defines the corresponding shower time $t$. There is a function $\lambda_{ll}(\{p,f\}_m, \zeta)$ that contains a factor $\as$ and depends on the splitting variables and the momenta and flavors of the partons before the splitting \cite{NSColorTheory}. This function contains a color factor $C_\LF$, $C_\LA$, or $T_\LR$ that is determined by the flavor of parton $l$. However, this factor is independent of the color state $\ket{\{c\}_m}$ on which it acts. That is, $v^\LCP_\textrm{coll}$ is proportional to the unit operator $I_\Lc$ on the ket color space.

The operator $a^\LCP_\textrm{soft}$ corresponds to graphs in which a gluon is exchanged between two partons, $l$ and $k$. The operator $a^\LCP_\textrm{soft}(t,\{p,f\}_m)$ has the form, when acting on a color basis vector,
\begin{equation}
  \label{eq:asoftdef}
  \begin{split}
  \frac{\as(t)}{2\pi}\,&
   a^\LCP_\textrm{soft}(t,\{p,f\}_m) \ket{\{c\}_m}
    \\
    &= 
    \sum_{l}\sum_{k \ne l}\int d\zeta\,
    \delta\big(t - T_l(\zeta,\{p\}_{m})\big)
    \\
    &\qquad\quad\times
    \lambda_{lk}(\{p,f\}_m, \zeta)\,\chi(k,l,\{c\}_m)
    \\&\times
    \ket{\{c\}_m}
    \;.
  \end{split}
\end{equation}
There is a function $\lambda_{lk}(\{p,f\}_m, \zeta)$ that contains a factor $\as$ and depends on the splitting variables and the momenta and flavors of the partons before the splitting.  Finally, the color information is provided by a function $\chi(k,l,\{c\}_m)$, which is 1 if partons $l$ and $k$ are next to each other along a color string in the basis state $\ket{\{c\}_m}$ and is 0 otherwise \cite{NSColorTheory}:
\begin{equation}
\begin{split}
\chi(k,l,&\{c\}_m) 
\\ ={}&
\begin{cases}
1 & k,l \textrm{ color connected in } \{c\}_m \\
0 & \textrm{otherwise}
\end{cases}
\;.
\end{split}
\end{equation}
Notice that the color basis states $\ket{\{c\}_m}$ are eigenvectors of $a^\LCP_\textrm{soft} (t, \{p,f\}_m)$.

The color basis states $\ket{\{c\}_m}$ are not eigenvectors of $T_\La\cdot T_\Lb$, so it would seem difficult to include the imaginary term proportional to $T_\La\cdot T_\Lb$ in the exponent in Eq.~(\ref{eq:noperator}). To see what is involved, let us start by considering the case that both incoming partons are antiquarks. Then incoming parton ``a'' carries an outgoing quark color index that we can call $\alpha$ and incoming parton ``b'' carries an outgoing quark color index that we can call $\beta$. The outgoing partons carry color indices that we can collectively denote by $\{r\}$. This gives us a (possible unnormalized) color basis state $\ket{G}$ with color wave function $G(\alpha,\beta,\{r\})$. With this notation, we can define what $T_\La\cdot T_\Lb$ does to $\ket{G}$:
\begin{equation}
T_\La\cdot T_\Lb \ket{G} = \ket{\widehat G}
\;,
\end{equation}
where, using the definition that $T_\La$ inserts a color generator on line ``a'' and $T_\Lb$ inserts a color generator on line ``a,''
\begin{equation}
\widehat G(\alpha,\beta;\{r\}) =
t(c)^{\alpha}_{\alpha''}t(c)^{\beta}_{\beta''}
G(\alpha'',\beta'';\{r\})
\;.
\end{equation}
There are implied sums over repeated color indices.

We will want to consider two basis states $\ket{G(n)}$, $n \in \{1,2\}$. We take $\ket{G(1)}$ to be the color basis state with which we start: $\ket{G(1)} = \ket{\{c\}_m}$. Then we define a second basis state $\ket{G(2)}$. For the color wave functions $G(n,\alpha,\beta,\{r\})$, we choose
\begin{equation}
G(n;\alpha,\beta;\{r\}) =  
R(\{r\})^{\alpha' \beta'} C(n;\alpha,\beta)_{\alpha' \beta'}
\;,
\end{equation}
where $R$ is a fixed color state vector that carries indices $\alpha', \beta'$ for two outgoing quark lines together with the indices $\{r\}$ for the rest of the partons. The two color wave functions $C(n;\alpha,\beta)_{\alpha' \beta'}$ are
\begin{equation}
\begin{split}
C(1;\alpha,\beta)_{\alpha' \beta'} ={}& 
\delta^{\alpha}_{\alpha'}\delta^{\beta}_{\beta'}
\;,
\\
C(2;\alpha,\beta)_{\alpha' \beta'} ={}& 
\delta^{\alpha}_{\beta'}\delta^{\beta}_{\alpha'}
\;.
\end{split}
\end{equation}

Then after applying $T_\La\cdot T_\Lb$, we get new basis states
\begin{equation}
\widehat G(n;\alpha,\beta;\{r\}) =  
R(\{r\})^{\alpha' \beta'} \widehat C(n;\alpha,\beta)_{\alpha' \beta'}
\;,
\end{equation}
where
\begin{equation}
\widehat C(n;\alpha,\beta)_{\alpha' \beta'} =
t(c)^{\alpha}_{\alpha''}t(c)^{\beta}_{\beta''}
C(n;\alpha'',\beta'')_{\alpha' \beta'}
\;.
\end{equation}

We can evaluate this. We can use the Fierz identity,
\begin{equation}
\label{eq:fierz}
t(c)^{\alpha}_{\alpha''}t(c)^{\beta}_{\beta''}
= \frac{1}{2}\,\delta^{\alpha}_{\beta''}\delta^{\beta}_{\alpha''}
-\frac{1}{2N_\Lc}\,\delta^{\alpha}_{\alpha''}\delta^{\beta}_{\beta''}
\;.
\end{equation}
This gives us
\begin{equation}
\begin{split}
\widehat C(n;\alpha,\beta)_{\alpha' \beta'} ={}&
\frac{1}{2}\,
C(n;\beta,\alpha)_{\alpha' \beta'}
\\&
-
\frac{1}{2N_\Lc}\,
C(n;\alpha,\beta)_{\alpha' \beta'}
\;.
\end{split}
\end{equation}
This calculation gives us
\begin{equation}
\widehat C(n;\alpha,\beta)_{\alpha'\beta'}
= \sum_{n'} M_{n'n}\,
C(n';\alpha,\beta)_{\alpha'\beta'}
\;,
\end{equation}
where
\begin{equation}
\label{eq:Mqbarqbar}
M = -\frac12 \left[
\begin{array}{cc}
    1/N_\Lc  & -1      \\
    -1       & 1/N_\Lc \\
\end{array}
\right]
\;.
\end{equation}
For our complete basis states $\ket{G(n)}$, this is
\begin{equation}
\label{eq:Mmatrix}
T_\La\cdot T_\Lb\ket{G(n)}
= \sum_{n'} M_{n'n} 
\ket{G(n')}
\;.
\end{equation}
That is, for two antiquark incoming states, the operator $T_\La\cdot T_\Lb$ acts as a matrix on just a two dimensional subspace of the color ket space or the color bra space. We do not need to deal with a color space of very high dimensionality.

Now we can return to the operator $n$ in Eq.~(\ref{eq:noperator}). The time ordered exponential in Eq.~(\ref{eq:noperator}) can be implemented as an ordinary exponential of a matrix by using the Magnus expansion \cite{Magnus:1954zz}. We have
\begin{equation}
\begin{split}
\label{eq:nexpansion}
n(t_2,{}& t_1,\{p,f\}_m) 
\\&=
\exp\!\left[-\sum_{k=1}^\infty \omega_k(t_2, t_1, \{p,f\}_m)\right]
\;.
\end{split}
\end{equation}
The first term in the exponent is simply $[\as(\tau)/(2\pi)]\, a(\tau,\{p,f\}_m)$ without time ordering:
\begin{equation}
\omega_1(t_2, t_1, \{p,f\}_m) = \int_{t_1}^{t_2} 
d\tau\ \frac{\as(\tau)}{2\pi}\, a(\tau,\{p,f\}_m)
\;.
\end{equation}
The second term in the exponent is
\begin{equation}
\begin{split}
\omega_2(t_2,{}& t_1, \{p,f\}_m) 
\\&= 
\int_{t_1}^{t_2} d\tau_1\frac{\as(\tau_1)}{2\pi}\int_{t_1}^{\tau_1} d\tau_2\, 
\frac{\as(\tau_2)}{2\pi}
\\&\qquad\times 
\big[a(\tau_1, \{p,f\}_m),\, a(\tau_2, \{p,f\}_m)\big]
\;.
\end{split}
\end{equation}
The higher order contributions can be obtained by recursion. 

The commutator in the second order term simplifies when we use the decomposition (\ref{eq:adecomposition}) of $a$. The operator $a^\LCP_\mathrm{coll}(t,\{p,f\}_m)$ contains the soft$\times$collinear singularities of $a$. Thus at fixed $t$, this operator contains a factor of $t$, which comes from integrating over the momentum fraction in a splitting. However, we saw in Eq.~(\ref{eq:acolldef}) that $a^\LCP_\mathrm{coll}(t,\{p,f\}_m)$ is proportional to the unit matrix, so that it commutes with the other terms in $a$. We also saw in Eq.~(\ref{eq:asoftdef}) that the soft part of $a$ is a diagonal operator. However, $T_\La\cdot T_\Lb$ does not commute with $a^\LCP_\mathrm{soft}$. Thus the commutator is 
\begin{align}
\label{eq:commutator} 
\big[{}&a(\tau_1,\{p,f\}_m),\, a(\tau_2,\{p,f\}_m)\big] 
\\
&= 
- 4 \mi \pi
\big[a^\LCP_\mathrm{soft}(\tau_1,\{p,f\}_m)
- a^\LCP_\mathrm{soft}(\tau_2,\{p,f\}_m), T_\La\cdot T_\Lb\big]
.
\notag
\end{align}
The operator $a^\LCP_\mathrm{soft}(t,\{p,f\}_m)$ contains a soft singularity only, so it does not have a contribution proportional to $t$. From its functional form \cite{NSColorTheory}, one sees that this operator is independent of the shower time except for shower times close to 0. It has the expansion
\begin{equation}
\begin{split}
a^\LCP_\mathrm{soft}(t, \{p,f\}_m) 
= {}&
a_\mathrm{soft}^{\LCP (0)}(\{p,f\}_m) 
\\& 
+ e^{-t} a_\mathrm{soft}^{\LCP (1)}(\{p,f\}_m) 
\\&+\cO(e^{-2t})
\;.
\end{split}
\end{equation}
With this, the commutator in Eq.~(\ref{eq:commutator}) is 
\begin{equation}
\label{eq:commutator1} 
\begin{split}
\big[{}&a(\tau_1, \{p,f\}_m),\, a(\tau_2,\{p,f\}_m)\big] 
\\
&= 
-4\mi\pi
(e^{-\tau_1}-e^{-\tau_2})\big[a_\mathrm{soft}^{\LCP (1)}(\{p,f\}_m), 
T_\La\cdot T_\Lb \big]
\\
&\quad + \cdots\;.
\end{split}
\end{equation}
We see that the second term of the exponent, $\omega_2$, in Eq.~(\ref{eq:nexpansion}) is not only second order in $\as$ but is also suppressed by powers of $e^{-t}$ so that it does not contribute in the soft and collinear enhanced regions. The higher order terms in Eq.~(\ref{eq:nexpansion}) are similarly suppressed. Thus these contributions can reasonably be neglected and we simply write $n$ in Eq.~(\ref{eq:noperator}) as the ordinary exponential of $a$.

We have seen that the operator $n$, including its $\mi\pi$ terms, can be written as an exponential of a finite dimensional matrix in the case of an incoming $\bar q$, $\bar q$ state. In the following subsections, we examine the other possible choices for the flavors of the initial state partons. In each case, we obtain a finite dimensional matrix that needs to be exponentiated numerically. The largest dimensionality is $14\times 14$.

\subsection{Incoming $q\, q$}

For an incoming $q\, q$ state, the incoming quarks carry outgoing antiquark indices. We then consider basis states of form
\begin{equation}
G(n;\alpha,\beta;\{r\}) =  
R(\{r\})_{\alpha' \beta'} C(n;\alpha,\beta)^{\alpha' \beta'}
\;,
\end{equation}
with
\begin{equation}
\begin{split}
C(1;\alpha,\beta)^{\alpha' \beta'} ={}& 
\delta^{\alpha'}_{\alpha}\delta^{\beta'}_{\beta}
\;,
\\
C(2;\alpha,\beta)^{\alpha' \beta'} ={}& 
\delta^{\alpha'}_{\beta}\delta^{\beta'}_{\alpha}
\;.
\end{split}
\end{equation}
When we exchange a gluon with octet color index $c$ between the two outgoing antiquark lines, we have
\begin{equation}
\widehat C(n;\alpha,\beta)^{\alpha' \beta'} =
t(c)^{\alpha''}_{\alpha}t(c)^{\beta''}_{\beta}
C(n;\alpha'',\beta'')^{\alpha' \beta'}
\;.
\end{equation}
Here the SU(3) generator for the first outgoing antiquark line is $-t(c)^{\alpha''}_{\alpha}$ and the generator for the second antiquark line is $-t(c)^{\beta''}_{\beta}$. The two $-1$ factors cancel. After using the Fierz identity, we obtain
\begin{equation}
\widehat C(n;\alpha,\beta)^{\alpha'\beta'}
= \sum_{n'} M_{n'n} 
C(n';\alpha,\beta)^{\alpha'\beta'}
\;,
\end{equation}
with the same matrix as in the $\bar q, \bar q$ case:
\begin{equation}
\label{eq:Mqq}
M = -\frac12 \left[
\begin{array}{cc}
    1/N_\Lc  & -1      \\
    -1       & 1/N_\Lc \\
\end{array}
\right]
\;.
\end{equation}

\subsection{Incoming $q\, \bar q$}

Here we have one outgoing quark index $\alpha$ and one outgoing antiquark index $\beta$. We then consider basis states of the form
\begin{equation}
G(n;\alpha,\beta;\{r\}) =  
R(\{r\})^{\alpha'}_{\beta'} C(n;\alpha,\beta)_{\alpha'}^{\beta'}
\;,
\end{equation}
with
\begin{equation}
\begin{split}
C(1;\alpha,\beta)_{\alpha'}^{\beta'} ={}& 
\delta^{\alpha}_{\alpha'}\delta^{\beta'}_{\beta}
\;,
\\
C(2;\alpha,\beta)_{\alpha'}^{\beta'} ={}& 
\delta^{\alpha}_{\beta}\delta^{\beta'}_{\alpha'}
\;.
\end{split}
\end{equation}
When we exchange a gluon with octet color index $c$ between the two external lines, we have
\begin{equation}
\widehat C(n;\alpha,\beta)_{\alpha'}^{\beta'} =
- t(c)^{\alpha}_{\alpha''}t(c)^{\beta''}_{\beta}
C(n;\alpha'',\beta'')_{\alpha'}^{\beta'}
\;.
\end{equation}
Here the SU(3) generator for the outgoing quark line is $t(c)^{\alpha}_{\alpha''}$ and the generator for the antiquark line is $-t(c)^{\beta''}_{\beta}$. After using the Fierz identity (\ref{eq:fierz}), we obtain
\begin{equation}
\widehat C(n;\alpha,\beta)_{\alpha'}^{\beta'}
= \sum_{n'} M_{n'n} 
C(n';\alpha,\beta)_{\alpha'}^{\beta'}
\;,
\end{equation}
with the a new matrix:
\begin{equation}
\label{eq:Mqqbar}
M = -\frac12 \left[
\begin{array}{cc}
 -1/N_\Lc  & 0       \\
   1       & 2 C_\LF \\
\end{array}
\right]
\;.
\end{equation}

\subsection{Incoming $\bar q\, \Lg$}

Here we have one outgoing quark index $\alpha$ and one outgoing gluon index $a$. We then consider basis states $\ket{G(n)}$, $n \in \{1,\dots,4\}$, with $\ket{G(1)} = \ket{\{c\}_n}$ and with color wave functions of the form
\begin{equation}
G(n;\alpha,a;\{r\}) =  
R(\{r\})^{\alpha'\beta'}_{\beta} C(n;\alpha,a)_{\alpha' \beta'}^{\beta}
\;.
\end{equation}
The matrix $R$ representing the rest of the color vector is given a quark index $\beta'$ and an antiquark index $\beta$ instead of a gluon index because in the trace basis the external gluon  couples to a quark line. We define
\begin{equation}
\begin{split}
C(1;a,\alpha)^{\beta}_{\alpha'\beta'} ={}& 
\delta^\alpha_{\alpha'}\,t(a)^\beta_{\beta'}
\;,
\\
C(2;a,\alpha)^{\beta}_{\alpha'\beta'} ={}& 
\delta^\alpha_{\beta'}\,t(a)^\beta_{\alpha'}
\;,
\\
C(3;a,\alpha)^{\beta}_{\alpha'\beta'} ={}& 
 \delta^\beta_{\beta'}\,t(a)^\alpha_{\alpha'}
\;,
\\
C(4;a,\alpha)^{\beta}_{\alpha'\beta'} ={}& 
\delta^\beta_{\alpha'}\,t(a)^\alpha_{\beta'}
\;.
\end{split}
\end{equation}
We use four basis states because there are four ways to attach the indices.

Suppose now that we exchange a gluon with octet color index $c$ between the outgoing quark line and the outgoing gluon line, giving us

\begin{equation}
\label{eq:hatCqg}
\widehat C(n;a,\alpha)^{\beta}_{\alpha'\beta'} 
= \mi f(a,c,a')\, t(c)^\alpha_{\alpha''}\, 
C(n;a',\alpha'')^{\beta}_{\alpha'\beta'}
\;.
\end{equation}

Let us start with $C(1;a,\alpha)^{\beta}_{\alpha'\beta'}$. We use the commutator rule, 
\begin{equation}
\label{eq:commutatorrule}
\mi f(a,c,a')\,t(a')^\alpha_{\alpha'}
= t(a)^{\alpha}_{\gamma}\,t(c)^{\gamma}_{\alpha'}
- t(c)^{\alpha}_{\gamma}\,t(a)^{\gamma}_{\alpha'}
\;.
\end{equation}
Then we use the first term of the Fierz identity (\ref{eq:fierz}), noting that the second term of the Fierz identity gives canceling contributions. This gives
\begin{equation}
\begin{split}
\widehat C(1;a,\alpha)^{\beta}_{\alpha'\beta'} ={}& 
\mi f(a,c,a')\, t(c)^\alpha_{\alpha''}\, \delta^{\alpha''}_{\alpha'}
t(a')^\beta_{\beta'}
\\
={}& \mi f(a,c,a')\, t(c)^\alpha_{\alpha'}\,
t(a')^\beta_{\beta'}
\\
={}& t(c)^\alpha_{\alpha'}\,
\{t(a)^\beta_\rho t(c)^\rho_{\beta'} - t(c)^\beta_\rho t(a)^\rho_{\beta'}\}
\\
={}& \frac{1}{2}\,
\{t(a)^\beta_\rho \delta^\rho_{\alpha'} \delta^\alpha_{\beta'} 
- t(a)^\rho_{\beta'}\delta^\beta_{\alpha'} \delta^\alpha_{\rho}\}
\\
={}& \frac{1}{2}\,
\{t(a)^\beta_{\alpha'} \delta^\alpha_{\beta'} 
- t(a)^\alpha_{\beta'}\delta^\beta_{\alpha'} \}
\;.
\end{split}
\end{equation}
This is
\begin{equation}
\widehat C(1;a,\alpha)^{\beta}_{\alpha'\beta'} =
\frac{1}{2}\,C(2;a,\alpha)^{\beta}_{\alpha'\beta'}
-\frac{1}{2}\,C(4;a,\alpha)^{\beta}_{\alpha'\beta'}
\;.
\end{equation}

The analogous calculation for $C(2;a,\alpha)^{\beta}_{\alpha'\beta'}$ gives
\begin{equation}
\begin{split}
\widehat C(2;a,\alpha)^{\beta}_{\alpha'\beta'}
={}& \frac{1}{2}\,
\{t(a)^\beta_{\beta'} \delta^\alpha_{\alpha'} 
- t(a)^\alpha_{\alpha'}\delta^\beta_{\beta'} \}
\;.
\end{split}
\end{equation}
This is
\begin{equation}
\widehat C(2;a,\alpha)^{\beta}_{\alpha'\beta'} =
\frac{1}{2}\,C(1;a,\alpha)^{\beta}_{\alpha'\beta'}
-\frac{1}{2}\,C(3;a,\alpha)^{\beta}_{\alpha'\beta'}
\;.
\end{equation}

Now we can turn to $C(3;a,\alpha)^{\beta}_{\alpha'\beta'}$:
\begin{equation}
\begin{split}
\widehat C(3;a,\alpha)^{\beta}_{\alpha'\beta'} ={}& 
\mi f(a,c,a')\, t(c)^\alpha_{\alpha''}\, 
t(a')^{\alpha''}_{\alpha'}\, \delta^\beta_{\beta'}\\
={}& t(c)^\alpha_{\alpha''}\,\delta^\beta_{\beta'}
\{t(a)^{\alpha''}_\rho t(c)^\rho_{\alpha'}
- t(c)^{\alpha''}_\rho t(a)^\rho_{\alpha'} \}
\\
={}& 
\delta^\beta_{\beta'} \{t(a)^{\alpha''}_\rho 
\delta^\alpha_{\alpha'}\delta^\rho_{\alpha''}
- t(a)^\rho_{\alpha'}
\delta^{\alpha''}_{\alpha''}\delta^\alpha_{\rho}
\}
\\
={}& 
\frac{1}{2}\,\delta^\beta_{\beta'}
\{t(a)^{\alpha''}_{\alpha''} \delta^\alpha_{\alpha'}
- t(a)^{\alpha}_{\alpha'}  \delta^{\alpha''}_{\alpha''}\}
\\
={}& 
-\frac{N_\Lc}{2}\,
t(a)^{\alpha}_{\alpha'}\delta^\beta_{\beta'}
\;.
\end{split}
\end{equation}
This is 
\begin{equation}
\widehat C(3;a,\alpha)^{\beta}_{\alpha'\beta'} =
-\frac{N_\Lc}{2}\,C(3;a,\alpha)^{\beta}_{\alpha'\beta'}
\;.
\end{equation}

The analogous calculation for $C(4;a,\alpha)^{\beta}_{\alpha'\beta'}$ gives
\begin{equation}
\begin{split}
\widehat C(4;a,\alpha)^{\beta}_{\alpha'\beta'} ={}& 
-\frac{N_\Lc}{2}\,
t(a)^{\alpha}_{\beta'}\delta^\beta_{\alpha'}
\;.
\end{split}
\end{equation}
This is 
\begin{equation}
\widehat C(4;a,\alpha)^{\beta}_{\alpha'\beta'} =
-\frac{N_\Lc}{2}\,C(4;a,\alpha)^{\beta}_{\alpha'\beta'}
\;.
\end{equation}

This calculation gives us
\begin{equation}
\widehat C(n;a,\alpha)^{\beta}_{\alpha'\beta'}
= \sum_{n'} M_{n'n}\,
C(n';a,\alpha)^{\beta}_{\alpha'\beta'}
\;,
\end{equation}
where
\begin{equation}
\label{eq:Mqbarg}
  M = -\frac12 \left[
  \begin{array}{rrrr}
    0  & -1 & 0 & 0 \\
    -1 & 0  & 0 & 0\\
    0 & 1 & N_\Lc & 0 \\
    1 & 0 & 0 & N_\Lc \\ 
  \end{array}
  \right]
  \;.
\end{equation}

\subsection{Incoming $q\, \Lg$}

In this case, we have one outgoing antiquark index $\alpha$ and one outgoing gluon index $a$. We then consider basis states of form
\begin{equation}
G(n;\alpha,a;\{r\}) =  
R(\{r\})^{\beta'}_{\alpha'\beta} C(n;\alpha,a)_{\beta'}^{\alpha'\beta}
\;.
\end{equation}
We define
\begin{equation}
\begin{split}
C(1;a,\alpha)_{\beta'}^{\alpha'\beta} ={}& 
\delta^{\alpha'}_{\alpha}\,t(a)^\beta_{\beta'}
\;,
\\
C(2;a,\alpha)_{\beta'}^{\alpha'\beta} ={}& 
\delta^{\beta}_{\alpha}\,t(a)^{\alpha'}_{\beta'}
\;,
\\
C(3;a,\alpha)_{\beta'}^{\alpha'\beta} ={}& 
\delta^{\beta}_{\beta'}\,t(a)^{\alpha'}_{\alpha}
\;,
\\
C(4;a,\alpha)_{\beta'}^{\alpha'\beta} ={}& 
\delta^{\alpha'}_{\beta'}\,t(a)^{\beta}_{\alpha}
\;.
\end{split}
\end{equation}

Suppose now that we exchange a gluon with octet color index $c$ between the outgoing antiquark line and the outgoing gluon line, giving us
\begin{equation}
\widehat C(n;a,\alpha)^{\alpha'\beta}_{\beta'} =
- t(c)^{\alpha''}_{\alpha}
\mi f_{aca'}
C(n;a', \alpha'')_{\beta'}^{\alpha'\beta}
\;.
\end{equation}
There is a minus sign here because we attach the color matrix to an antiquark line.

Then a calculation using the identities (\ref{eq:commutatorrule}) and (\ref{eq:fierz}) gives us
\begin{equation}
\widehat C(n;a,\alpha)^{\alpha'\beta}_{\beta'}
= \sum_{n'} M_{n'n} 
C(n';a,\alpha)^{\alpha'\beta}_{\beta'}
\;,
\end{equation}
where
\begin{equation}
\label{eq:Mqg}
M = -\frac12 \left[
\begin{array}{rrrr}
    0  & -1 & 0 & 0   \\
    -1 & 0  & 0 & 0    \\
    0 & 1 & N_\Lc & 0 \\
    1 & 0 & 0 & N_\Lc \\ 
\end{array}
\right]
\;.
\end{equation}

\subsection{Incoming $\Lg\, \Lg$}

In this case, we have two outgoing gluon indices $a$ and $b$. We consider basis states of form
\begin{equation}
G(n;a,b;\{r\}) =  
R(\{r\})^{\alpha'\beta'}_{\alpha\beta} C(n;a,b)^{\alpha\beta}_{\alpha'\beta'}
\;.
\end{equation}
Each vector $C(n;a,b)$ carries two (outgoing) quark indices $\alpha$ and $\beta$, and two (outgoing) antiquark indices $\alpha'$ and $\beta'$. There are 14 ways to couple the two gluon indices to the quark and antiquark indices:
\begin{equation}
\begin{split}
C(1;a,b)^{\alpha\beta}_{\alpha'\beta'} ={}& 
t(a)^\alpha_{\alpha'}\ t(b)^\beta_{\beta'}
\;,
\\
C(2;a,b)^{\alpha\beta}_{\alpha'\beta'} ={}& 
t(a)^\beta_{\beta'}\ t(b)^\alpha_{\alpha'}
\;,
\\
C(3;a,b)^{\alpha\beta}_{\alpha'\beta'} ={}& 
t(a)^\alpha_{\beta'}\ t(b)^\beta_{\alpha'}
\;,
\\
C(4;a,b)^{\alpha\beta}_{\alpha'\beta'} ={}& 
t(a)^\beta_{\alpha'}\ t(b)^\alpha_{\beta'}
\;,
\\
C(5;a,b)^{\alpha\beta}_{\alpha'\beta'} 
={}& t(a)^\alpha_{\gamma}\, t(b)^\gamma_{\beta'}\ \delta^\beta_{\alpha'}
\;,
\\
C(6;a,b)^{\alpha\beta}_{\alpha'\beta'} ={}& 
t(b)^\beta_{\gamma}\, t(a)^\gamma_{\alpha'}\ \delta^\alpha_{\beta'}
\;,
\\
C(7;a,b)^{\alpha\beta}_{\alpha'\beta'} ={}& 
t(b)^\alpha_{\gamma}\, t(a)^\gamma_{\beta'}\ \delta^\beta_{\alpha'}
\;,
\\
C(8;a,b)^{\alpha\beta}_{\alpha'\beta'} ={}& 
t(a)^\beta_{\gamma}\, t(b)^\gamma_{\alpha'}\ \delta^\alpha_{\beta'}
\;,
\\
C(9;a,b)^{\alpha\beta}_{\alpha'\beta'} ={}& 
t(a)^\alpha_{\gamma}\, t(b)^\gamma_{\alpha'}\ \delta^\beta_{\beta'}
\;,
\\
C(10;a,b)^{\alpha\beta}_{\alpha'\beta'} ={}& 
t(b)^\beta_{\gamma}\, t(a)^\gamma_{\beta'}\ \delta^\alpha_{\alpha'}
\;,
\\
C(11;a,b)^{\alpha\beta}_{\alpha'\beta'} ={}& 
t(b)^\alpha_{\gamma}\, t(a)^\gamma_{\alpha'}\ \delta^\beta_{\beta'}
\;,
\\
C(12;a,b)^{\alpha\beta}_{\alpha'\beta'} ={}& 
t(a)^\beta_{\gamma}\, t(b)^\gamma_{\beta'}\ \delta^\alpha_{\alpha'}
\;,
\\
C(13;a,b)^{\alpha\beta}_{\alpha'\beta'} 
={}& t(a)^\gamma_{\delta}\, t(b)^\delta_{\gamma}\ 
\delta^\alpha_{\beta'}\ \delta^\beta_{\alpha'}
\;,
\\
C(14;a,b)^{\alpha\beta}_{\alpha'\beta'} 
={}& t(a)^\gamma_{\delta}\, t(b)^\delta_{\gamma}\ 
\delta^\alpha_{\alpha'}\ \delta^\beta_{\beta'}
\;.
\end{split}
\end{equation}

Suppose now that we exchange a gluon with octet color index $c$ between the outgoing gluon lines, giving us
\begin{equation}
\widehat C(n;a,b)^{\alpha\beta}_{\alpha'\beta'} =
\mi f_{aca'}
\mi f_{bcb'}
C(n;a',b')^{\alpha\beta}_{\alpha'\beta'}
\;.
\end{equation}
Then a calculation similar to the calculations that we have done previously gives
\begin{equation}
\widehat C(n;a,b)^{\alpha\beta}_{\alpha'\beta'}
= \sum_{n'} M_{n'n} 
C(n';a,b)^{\alpha\beta}_{\alpha'\beta'}
\;,
\end{equation}
where
\begin{widetext} 
\begin{equation}
M = -\frac12 \left[
  \begin{array}{rrrrrrrrrrrrrr}
    0 & 0 &-1 &-1 & 0 & 0 & 0 & 0 & 0 & 0  & 0  & 0  &  0 & 0  \\
    0 & 0 &-1 &-1 & 0 & 0 & 0 & 0 & 0 & 0  & 0  & 0  &  0 & 0  \\
   -1 &-1 & 0 & 0 & 0 & 0 & 0 & 0 & 0 & 0  & 0  & 0  &  0 & 0  \\
   -1 &-1 & 0 & 0 & 0 & 0 & 0 & 0 & 0 & 0  & 0  & 0  &  0 & 0  \\
    1 & 0 & 0 & 0 & N_c & 0 & 0 & 0 & 0 & 0  & 0  & 0  &  0 & 0  \\   
    1 & 0 & 0 & 0 & 0 & N_c & 0 & 0 & 0 & 0  & 0  & 0  &  0 & 0  \\   
    0 & 1 & 0 & 0 & 0 & 0 & N_c & 0 & 0 & 0  & 0  & 0  &  0 & 0  \\   
    0 & 1 & 0 & 0 & 0 & 0 & 0 & N_c & 0 & 0  & 0  & 0  &  0 & 0  \\   
    0 & 0 & 1 & 0 & 0 & 0 & 0 & 0 & N_c & 0  & 0  & 0  &  0 & 0  \\   
    0 & 0 & 1 & 0 & 0 & 0 & 0 & 0 & 0 & N_c  & 0  & 0  &  0 & 0  \\   
    0 & 0 & 0 & 1 & 0 & 0 & 0 & 0 & 0 & 0  & N_c  & 0  &  0 & 0  \\   
    0 & 0 & 0 & 1 & 0 & 0 & 0 & 0 & 0 & 0  & 0  & N_c  &  0 & 0  \\   
    0 & 0 & 0 & 0 & 1 & 1 & 1 & 1 & 0 & 0  & 0  & 0  &  2N_c & 0  \\   
    0 & 0 & 0 & 0 & 0 & 0 & 0 & 0 & 1 & 1  & 1  & 1  &  0 & 2N_c  \\   
  \end{array}
  \right]
\;.
\end{equation}

\end{widetext} 

\section{\label{sec:assembling}Assembling the shower}

It is now clear what to do to complete the shower evolution equations. Starting with states $\ket{G(1)} = \ket{\{c\}_m}$ and $\bra{G'(1)} = \bra{\{c'\}_m}$, we define color states $\ket{G(n)}$ and $\bra{G'(n)}$. This gives us an operator $n(t_2,t_1,\{p,f\}_m)$ defined as a matrix that mixes the states $\ket{G(n)}$ and an operator $n^\dagger(t_2,t_1,\{p,f\}_m)$ that mixes the states $\bra{G'(n)}$ according to Eqs.~(\ref{eq:noperator}), (\ref{eq:adecomposition}), and (\ref{eq:Mmatrix}). Then the combined no-splitting operator $\cN^\scE(t_2,t_1)$ is defined by the direct product of these matrices, according to Eqs.~(\ref{eq:Nmatrix}),  (\ref{eq:Nton}), and (\ref{eq:Ntonstar}).

Now, in a complete algorithm, one would calculate $\cX^\scE(t_2,t_2)$ as a power series in $\Delta \cV_\textrm{Re}(\tau)$ by using Eq.~(\ref{eq:evolutionXnew}). Then one would produce the complete shower using Eq.~(\ref{eq:evolutionU6}). One would keep any number of splittings generated by $\cH^\LCP(\tau)$ and $\Delta \cH(\tau)$. This would generate contributions proportional to powers of $\Delta \cH(\tau)$ and $\Delta \cV_\textrm{Re}(\tau)$. One would retain contributions proportional to $[\Delta \cH]^A  [\Delta \cV_\textrm{Re}]^B$ with $A + B \le N_\textrm{Re}$ for a chosen value of $N_\textrm{Re}$.

For this paper, we have not implemented the complete algorithm. Rather, we have set $N_\textrm{Re} = 0$. This allows us to test whether the exponentiation of the $\mi\pi$ contributions as outlined above is computationally practical. It also allows us to test whether exponentiation of the $\mi\pi$ contributions makes a substantial numerical difference compared to simply keeping just a small number of factors of $\cV_{\mi\pi}$. For this purpose, we will use the gap fraction, for which we have found previously that the inclusion of color beyond the LC+ approximation can be numerically important \cite{NSGapColor}.

\section{\label{sec:results}Results}

In this section, we explore the exponentiation of $\cV_{\mi\pi}$ by examining its effect on the gap survival probability for jet production in proton-proton collisions at $\sqrt s = 13 \TeV$. Using the anti-$k_\LT$ jet algorithm \cite{antikt} with a radius parameter $R = 0.4$, we find jets with transverse momenta $P_{\LT}$ and rapidities $y$ in the region $-Y_\mathrm{cut} < y < Y_\mathrm{cut}$. We will use $Y_\mathrm{cut} = 4.4$. Label the two highest $P_\LT$ jets 1 and 2, with $y_1 < y_2$. Define
\begin{equation}
\begin{split}
\bar p_\LT ={}& \frac{1}{2} (P_{\LT,1} + P_{\LT,2})
\;,
\\
y_{12} ={}& y_2 - y_1
\;.
\end{split}
\end{equation}
Now, define a cut parameter $p_{\LT}^\mathrm{cut}$. We will take $p_{\LT}^\mathrm{cut} = 20 \GeV$. Look at those jets with $P_\LT >  p_{\LT}^\mathrm{cut}$ in the rapidity region $y_1 < y < y_2$ between the two leading jets. We will say that the event has a rapidity gap if there are no such jets in this rapidity region. We denote the fraction of events with a rapidity gap by $f(\bar p_{\LT}, y_{12})$.

As discussed in the previous section, we have not implemented the inclusion of the operators $\Delta \cH(\tau)$ and $\Delta \cV_\textrm{Re}(\tau)$ in the shower along with the exponentiation of $ \cV_{\mi\pi}(\tau)$. For that reason, we do not include $\Delta \cH(\tau)$ and $\Delta \cV_\textrm{Re}(\tau)$ in the shower operator $\cU(t,t_0)$ in any of the calculations in this section.

Just after the hard scattering, \textsc{Deductor} inserts an operator $\cU_\cV$ that produces a summation of threshold logarithms \cite{NSThresholdII}. The operator $\cU_\cV$ gives results as an expansion in powers of $\Delta \cV_\mathrm{Re}$. \textsc{Deductor} retains only those terms with no more than $N_\Delta^\mathrm{thr}$ factors of $\Delta \cV_\mathrm{Re}$. We choose $N_\Delta^\mathrm{thr} = 0$ or $N_\Delta^\mathrm{thr} = 1$, as we will specify, in the investigations in this section.

We begin by examining results for $f(\bar p_{\LT}, y_{12})$ with several values of $N_{\mi\pi}$ and with the $\cV_{\mi\pi}$ contributions exponentiated. Our aim is to see if the perturbative results with increasing values of $N_{\mi\pi}$ are consistent with the exponentiated result. In this investigation, we examine $f$ in the range $300 \GeV < \bar p_\LT < 400 \GeV$ and $4 < y_{12} < 5$. We let the parton shower run over the evolution range $\mu_\Ls > \Lambda > 30 \GeV$ and stop evolution at $\Lambda = 30 \GeV$. Here $\mu_\Ls = 3 \bar p_\LT/2$, as in Ref.~\cite{NSGapColor}.

Later in this section, we will choose the maximum color suppression index $I_\mathrm{max} = 4$. However, this choice allows a systematic error of about $1/N_\Lc^4 \approx 0.01$ and we want to avoid that here, so we choose $I_\mathrm{max} = 6$, even though this gives us larger statistical errors.

\begin{figure}
\begin{center}
\ifusefigs 

\begin{tikzpicture}
  \begin{axis}[title = {Effect of $N_{\mi\pi}$},
    xlabel={$N_{\mi\pi}$}, ylabel={$f$},
    ymin=0.16, ymax=0.24,
    ytick={0.16,0.18,0.20,0.22,0.24},
    yticklabels={0.16,0.18,0.20,0.22,0.24},
    xtick = {0,2,4,6,8,10},
    xticklabels = {0,2,4,6,8,$\infty$},
    ]
    
\addplot +[only marks, error bars/.cd, y dir = both, y explicit]
coordinates{
(0.0,0.201483)   +- (0,0.000949353) 
(2.0,0.215155)   +- (0,0.00175589)
(4.0,0.211753)   +- (0,0.0019833)
(6.0,0.216401)   +- (0,0.00501215)
(8.0,0.199948)   +- (0,0.00433574)
};

\addplot +[red, only marks, error bars/.cd, y dir = both, y explicit]
coordinates{
(10.0,0.204467792)   +- (0,0.004339)
};

\end{axis}
\end{tikzpicture}

\else 
\NOTE{Figure fig:NiPi goes here.}
\fi

\end{center}
\caption{
Gap fraction $f$ calculated with different values of $N_{\mi\pi}$ with $N_\mathrm{Re} = N_\Delta^\mathrm{thr} = 0$. The red point shown with the label  $N_{\mi \pi} = \infty$  is the exponentiated version. We use maximum color suppression index $I_\textrm{max} = 6$.
}
\label{fig:NiPi}
\end{figure}
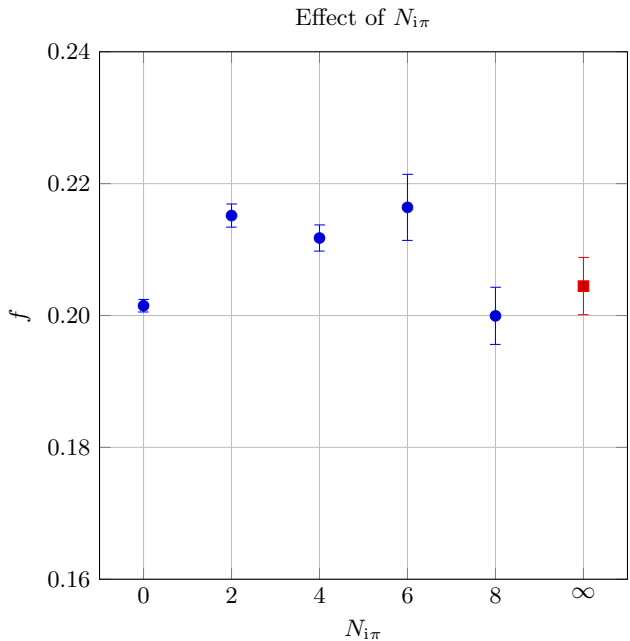

In Fig.~\ref{fig:NiPi}, we exhibit the results for $f$ with $N_{\mi\pi} = 0,2,4,6,8$ and then with the $\cV_{\mi\pi}$ contributions exponentiated, labelled $N_{\mi\pi} = \infty$. For Fig.~\ref{fig:NiPi}, we choose $N_\Delta^\mathrm{thr} = 0$. We see that there is a jump of $0.014$ between $N_{\mi\pi} = 0$ and $N_{\mi\pi} = 2$. For larger values of $N_{\mi\pi}$, the results appear to oscillate, but the statistical errors increase. All that we can really say is that the result for large $N_{\mi\pi}$ seems to be between 0.200 and 0.215. This is consistent with the exponentiated result, $f = 0.204 \pm 0.04$. That is, the perturbative results and the exponentiated result agree that the change from the $N_{\mi\pi} = 0$ result is $\Delta f \approx 0.003 \pm 0.004$. It is remarkable that the effect of $\cV_{\mi\pi}$ is so small. If the effect were of order $\Delta f \approx 0.1$, it would be easy to see, but an effect of order $\Delta f \approx 0.01$ is difficult to see in the face of systematic and statistical errors of the program.

In Ref.~\cite{NSGapColor}, we examined the effect of color on the rapidity gap survival fraction for a range of $y_{12}$ and $\bar p_\LT$. We found that the effect of having up to two powers of $\cV_{\mi\pi}$ was always quite small. It is of interest to see if the effect of including $\cV_{\mi\pi}$ in exponentiated form is also small for all $y_{12}$ and $\bar p_\LT$. 

For this study, results were obtained with the stated values $N_{\mi\pi}$ or with exponentiated $\cV_{\mi\pi}$ in the shower between $\Lambda = \mu_\Ls$ and $\Lambda = \Lambda_\mathrm{min} = 30 \GeV$. For the rest of the shower, down to $\Lambda = 1 \GeV$, we used the LC+ approximation.  Throughout, we set the maximum color suppression index to $I_\mathrm{max} = 4$ and we set $N_\Delta^\mathrm{thr} = 1$.

\begin{figure}
\begin{center}
\ifusefigs 

\else 
\NOTE{Figure fig:gap goes here.}
\fi

\begin{tikzpicture}
 \begin{groupplot}[
      group style={
          group size=1 by 5,
          vertical sep=0pt,
          x descriptions at=edge bottom},
          xlabel={$\bar p_\LT\,\mathrm{[GeV]}$},
          width=8.6cm,
    ]


\nextgroupplot[
ylabel={$f(\bar p_\LT)$},
height=5cm,
xmin=50, xmax=500,
ymin=0.1, ymax=1.0,
xmode=log,
ymode=log,
ytick={0.1,0.2,0.3,0.4,0.5,0.6,0.7,0.8,0.9,1.0},
yticklabels={,0.2,0.3,0.4,,0.6,,0.8,,1.0},
xtick={50,60,70,80,90,100,200,300,400,500},
  legend cell align=left,
  every axis legend/.append style={
  at={(0.03,0.05)},
  anchor = south west},
]

\pgfplotstableread{
 5.500000e+01     6.759508e-01   8.024205e-01   6.895477e-01   8.164659e-01 	      6.865641e-01   8.079672e-01
 6.500000e+01     6.754357e-01   8.417969e-01   6.904311e-01   8.563396e-01 	      6.645018e-01   8.311305e-01
 7.500000e+01     6.621223e-01   8.657893e-01   6.801280e-01   8.831818e-01 	      6.446468e-01   8.398050e-01
 8.500000e+01     6.431786e-01   8.720834e-01   6.678941e-01   8.930330e-01 	      6.448984e-01   8.719209e-01
 9.750000e+01     6.264260e-01   8.951918e-01   6.575420e-01   9.280826e-01 	      6.301845e-01   8.946747e-01
 1.125000e+02     6.140194e-01   9.153975e-01   6.405559e-01   9.483630e-01 	      6.205242e-01   9.218788e-01
 1.275000e+02     5.850775e-01   9.146435e-01   6.275327e-01   9.565304e-01 	      5.879088e-01   9.083444e-01
 1.425000e+02     5.688342e-01   9.230148e-01   5.984182e-01   9.480843e-01 	      5.863105e-01   9.465022e-01
 1.650000e+02     5.516540e-01   9.440784e-01   5.879447e-01   9.810277e-01 	      5.578339e-01   9.509952e-01
 1.950000e+02     5.261800e-01   9.513576e-01   5.599960e-01   9.838028e-01 	      5.380412e-01   9.607695e-01
 2.250000e+02     5.001390e-01   9.538980e-01   5.282008e-01   9.777388e-01 	      5.165443e-01   9.658487e-01
 2.550000e+02     4.842681e-01   9.619566e-01   5.140844e-01   9.838566e-01 	      4.909391e-01   9.564861e-01
 2.850000e+02     4.710700e-01   9.651787e-01   4.950714e-01   9.851062e-01 	      4.773122e-01   9.775381e-01
 3.200000e+02     4.540856e-01   9.678875e-01   4.801695e-01   9.864787e-01 	      4.624770e-01   9.512617e-01
 3.600000e+02     4.414227e-01   9.749785e-01   4.690033e-01   9.987709e-01 	      4.456635e-01   9.776871e-01
 4.000000e+02     4.271864e-01   9.697472e-01   4.550199e-01   9.936847e-01 	      4.411145e-01   9.794093e-01
 4.400000e+02     4.187534e-01   9.746573e-01   4.456944e-01   1.001814e+00 	      4.097025e-01   9.612637e-01
 4.800000e+02     4.046703e-01   9.736884e-01   4.280797e-01   9.946953e-01 	      4.025048e-01   9.485522e-01
 }\ductA

\addplot [blue,semithick] table [x={0},y expr=\thisrow{1}/\thisrow{2}]{\ductA};
\addlegendentry{$N_{\mathrm{i}\pi} = 0$}

\addplot [darkgreen,semithick] table [x={0},y expr=\thisrow{3}/\thisrow{4}]{\ductA};
\addlegendentry{$N_{\mathrm{i}\pi} = 2$}

\addplot [red,semithick] table [x={0},y expr=\thisrow{5}/\thisrow{6}]{\ductA};
\addlegendentry{$N_{\mathrm{i}\pi} = \infty$}

\node at (350.0,0.8) {\fcolorbox{black}{white}{\small{$1 < y_{12} < 2$}}};


\nextgroupplot[
ylabel = $f(\bar p_\LT)$,
height = 5 cm,
xmin=50, xmax=500,
ymin=0.1, ymax=1.0,
xmode=log,
ymode=log,
ytick={0.1,0.2,0.3,0.4,0.5,0.6,0.7,0.8,0.9,1.0},
yticklabels={,0.2,0.3,0.4,,0.6,,0.8,,1.0},
xtick={50,60,70,80,90,100,200,300,400,500},
]

\pgfplotstableread{
 5.500000e+01     6.151489e-01   8.019047e-01   6.377996e-01   8.259793e-01 	      6.183191e-01   8.030362e-01
 6.500000e+01     5.908621e-01   8.299817e-01   6.194986e-01   8.613219e-01 	      5.960672e-01   8.342650e-01
 7.500000e+01     5.754228e-01   8.631411e-01   6.129160e-01   9.025861e-01 	      5.772291e-01   8.519212e-01
 8.500000e+01     5.525974e-01   8.755260e-01   5.934249e-01   9.207373e-01 	      5.575270e-01   8.836018e-01
 9.750000e+01     5.237647e-01   8.961771e-01   5.680897e-01   9.403529e-01 	      5.227132e-01   8.906627e-01
 1.125000e+02     4.872798e-01   9.037761e-01   5.283642e-01   9.477046e-01 	      5.007248e-01   9.052671e-01
 1.275000e+02     4.709639e-01   9.229699e-01   5.167713e-01   9.730104e-01 	      4.698796e-01   8.878750e-01
 1.425000e+02     4.453194e-01   9.255821e-01   4.823782e-01   9.613011e-01 	      4.477647e-01   9.207599e-01
 1.650000e+02     4.223976e-01   9.432093e-01   4.655505e-01   9.856364e-01 	      4.150932e-01   9.268452e-01
 1.950000e+02     3.871050e-01   9.544756e-01   4.254024e-01   9.935861e-01 	      3.932686e-01   9.476057e-01
 2.250000e+02     3.645696e-01   9.508368e-01   3.999392e-01   9.807511e-01 	      3.842554e-01   9.571747e-01
 2.550000e+02     3.414416e-01   9.500433e-01   3.765698e-01   9.841185e-01 	      3.409991e-01   9.481959e-01
 2.850000e+02     3.311643e-01   9.576070e-01   3.592158e-01   9.847830e-01 	      3.436380e-01   9.713205e-01
 3.200000e+02     3.152775e-01   9.644510e-01   3.456306e-01   9.951739e-01 	      3.236888e-01   9.668360e-01
 3.600000e+02     3.014505e-01   9.605291e-01   3.274538e-01   9.791894e-01 	      3.098293e-01   9.615513e-01
 4.000000e+02     2.873260e-01   9.634554e-01   3.152127e-01   9.855925e-01 	      2.968425e-01   9.520008e-01
 4.400000e+02     2.780481e-01   9.648851e-01   3.058662e-01   9.860058e-01 	      2.915375e-01   9.509098e-01
 4.800000e+02     2.732208e-01   9.667399e-01   2.994770e-01   9.885552e-01 	      2.792480e-01   9.492762e-01
}\ductB

\addplot [blue,semithick] table [x={0},y expr=\thisrow{1}/\thisrow{2}]{\ductB};
\addplot [darkgreen,semithick] table [x={0},y expr=\thisrow{3}/\thisrow{4}]{\ductB};
\addplot [red,semithick] table [x={0},y expr=\thisrow{5}/\thisrow{6}]{\ductB};

\node at (350.0,0.8) {\fcolorbox{black}{white}{\small{$2 < y_{12} < 3$}}};


\nextgroupplot[
ylabel = $f(\bar p_\LT)$,
height = 5 cm,
xmin=50, xmax=500,
ymin=0.1, ymax=1.0,
xmode=log,
ymode=log,
ytick={0.1,0.2,0.3,0.4,0.5,0.6,0.7,0.8,0.9,1.0},
yticklabels={,0.2,0.3,0.4,,0.6,,0.8,,1.0},
xtick={50,60,70,80,90,100,200,300,400,500},]

\pgfplotstableread{
 5.500000e+01     5.487704e-01   7.893958e-01   5.788396e-01   8.241866e-01 	      5.436518e-01   7.760967e-01
 6.500000e+01     5.152562e-01   8.193158e-01   5.503101e-01   8.577739e-01 	      5.265449e-01   8.304195e-01
 7.500000e+01     4.925191e-01   8.532355e-01   5.328241e-01   8.994052e-01 	      4.946036e-01   8.435596e-01
 8.500000e+01     4.637922e-01   8.643315e-01   5.063624e-01   9.173248e-01 	      4.745455e-01   8.594767e-01
 9.750000e+01     4.431029e-01   8.980557e-01   4.840326e-01   9.417539e-01 	      4.406341e-01   8.899107e-01
 1.125000e+02     4.134618e-01   9.089901e-01   4.519471e-01   9.458278e-01 	      4.193922e-01   8.932137e-01
 1.275000e+02     3.763591e-01   8.994376e-01   4.197250e-01   9.446408e-01 	      3.846723e-01   9.032770e-01
 1.425000e+02     3.663404e-01   9.361943e-01   4.083008e-01   9.769814e-01 	      3.706606e-01   9.138319e-01
 1.650000e+02     3.217322e-01   9.230503e-01   3.561956e-01   9.638129e-01 	      3.444050e-01   9.394101e-01
 1.950000e+02     2.961919e-01   9.289406e-01   3.296914e-01   9.622390e-01 	      3.036183e-01   9.175071e-01
 2.250000e+02     2.770534e-01   9.528449e-01   3.069635e-01   9.895690e-01 	      2.832177e-01   9.400700e-01
 2.550000e+02     2.563933e-01   9.429891e-01   2.853537e-01   9.748617e-01 	      2.677104e-01   9.223709e-01
 2.850000e+02     2.413038e-01   9.418848e-01   2.657497e-01   9.676519e-01 	      2.482612e-01   9.320069e-01
 3.200000e+02     2.311625e-01   9.446332e-01   2.580886e-01   9.726223e-01 	      2.379637e-01   9.417771e-01
 3.600000e+02     2.207563e-01   9.480842e-01   2.444573e-01   9.689595e-01 	      2.271111e-01   9.458695e-01
 4.000000e+02     2.073169e-01   9.518497e-01   2.345449e-01   9.748915e-01 	      2.174530e-01   9.568156e-01
 4.400000e+02     2.037043e-01   9.507215e-01   2.313043e-01   9.714302e-01 	      2.106294e-01   9.593869e-01
 4.800000e+02     1.974196e-01   9.568948e-01   2.256258e-01   9.822881e-01 	      2.048822e-01   9.510776e-01
}\ductC

\addplot [blue,semithick] table [x={0},y expr=\thisrow{1}/\thisrow{2}]{\ductC};
\addplot [darkgreen,semithick] table [x={0},y expr=\thisrow{3}/\thisrow{4}]{\ductC};
\addplot [red,semithick] table [x={0},y expr=\thisrow{5}/\thisrow{6}]{\ductC};

\node at (350.0,0.8) {\fcolorbox{black}{white}{\small{$3 < y_{12} < 4$}}};


\nextgroupplot[
ylabel = $f(\bar p_\LT)$,
height = 5 cm,
xmin=50, xmax=500,
ymin=0.1, ymax=1.0,
xmode=log,
ymode=log,
ytick={0.1,0.2,0.3,0.4,0.5,0.6,0.7,0.8,0.9,1.0},
yticklabels={,0.2,0.3,0.4,,0.6,,0.8,,1.0},
xtick={50,60,70,80,90,100,200,300,400,500},]

\pgfplotstableread{
 5.500000e+01     5.023084e-01   8.001773e-01   5.327808e-01   8.364325e-01 	      5.006222e-01   7.951688e-01
 6.500000e+01     4.591406e-01   8.285970e-01   4.956341e-01   8.711418e-01 	      4.757308e-01   8.166935e-01
 7.500000e+01     4.350248e-01   8.473789e-01   4.721947e-01   8.923361e-01 	      4.558894e-01   8.808697e-01
 8.500000e+01     4.096358e-01   8.691398e-01   4.449477e-01   9.143029e-01 	      4.151466e-01   8.497992e-01
 9.750000e+01     3.826973e-01   9.030113e-01   4.185577e-01   9.447034e-01 	      3.805925e-01   8.667904e-01
 1.125000e+02     3.485625e-01   8.871379e-01   3.802429e-01   9.296150e-01 	      3.581764e-01   9.255099e-01
 1.275000e+02     3.221930e-01   9.048716e-01   3.511869e-01   9.429276e-01 	      3.329639e-01   8.940359e-01
 1.425000e+02     2.980541e-01   9.050544e-01   3.234703e-01   9.223961e-01 	      3.018829e-01   8.981096e-01
 1.650000e+02     2.690549e-01   9.014147e-01   2.931767e-01   9.301653e-01 	      2.686352e-01   8.863045e-01
 1.950000e+02     2.456901e-01   9.029219e-01   2.712496e-01   9.250123e-01 	      2.415561e-01   8.830965e-01
 2.250000e+02     2.198447e-01   9.188758e-01   2.450208e-01   9.491516e-01 	      2.230602e-01   9.554712e-01
 2.550000e+02     2.142931e-01   9.251456e-01   2.357979e-01   9.464234e-01 	      2.085016e-01   9.242151e-01
 2.850000e+02     2.007660e-01   9.462466e-01   2.258248e-01   9.682635e-01 	      2.002414e-01   9.229618e-01
 3.200000e+02     1.956339e-01   9.410779e-01   2.187410e-01   9.628660e-01 	      1.918832e-01   9.095293e-01
 3.600000e+02     1.850466e-01   9.335404e-01   2.059681e-01   9.575668e-01 	      1.888005e-01   9.439098e-01
 4.000000e+02     1.787895e-01   9.385163e-01   2.028526e-01   9.589788e-01 	      1.800068e-01   9.548944e-01
 4.400000e+02     1.773023e-01   9.340411e-01   2.039184e-01   9.535294e-01 	      1.796094e-01   9.266601e-01
 4.800000e+02     1.698800e-01   9.264074e-01   1.934949e-01   9.507070e-01 	      1.747322e-01   9.463222e-01
}\ductD

\addplot [blue,semithick] table [x={0},y expr=\thisrow{1}/\thisrow{2}]{\ductD};
\addplot [darkgreen,semithick] table [x={0},y expr=\thisrow{3}/\thisrow{4}]{\ductD};
\addplot [red,semithick] table [x={0},y expr=\thisrow{5}/\thisrow{6}]{\ductD};

\node at (350.0,0.8) {\fcolorbox{black}{white}{\small{$4 < y_{12} < 5$}}};


\nextgroupplot[
ylabel={$f(\bar p_\LT)$},
height=5cm,
xmin=50, xmax=500,
ymin=0.1, ymax=1.0,
xmode=log,
ymode=log,
ytick={0.1,0.2,0.3,0.4,0.5,0.6,0.7,0.8,0.9,1.0},
yticklabels={0.1,0.2,0.3,0.4,,0.6,,0.8,,1.0},
xtick={50,60,70,80,90,100,200,300,400,500},
xticklabels={50,60,,80,,100,200,300,400,500},]

\pgfplotstableread{
 5.500000e+01     4.362751e-01   7.876233e-01   4.623564e-01   8.159999e-01 	      4.349608e-01   7.938186e-01
 6.500000e+01     4.067262e-01   8.189687e-01   4.348474e-01   8.545389e-01 	      4.074319e-01   8.126894e-01
 7.500000e+01     3.801429e-01   8.436026e-01   4.097122e-01   8.882576e-01 	      3.780276e-01   8.276762e-01
 8.500000e+01     3.565139e-01   8.642294e-01   3.873672e-01   9.012420e-01 	      3.603675e-01   8.585761e-01
 9.750000e+01     3.358156e-01   8.749660e-01   3.611629e-01   9.086183e-01 	      3.508913e-01   9.177836e-01
 1.125000e+02     3.173118e-01   9.056275e-01   3.399596e-01   9.393476e-01 	      3.211100e-01   9.318355e-01
 1.275000e+02     2.970201e-01   9.034467e-01   3.173999e-01   9.448435e-01 	      2.843586e-01   8.843243e-01
 1.425000e+02     2.696289e-01   8.759813e-01   2.886007e-01   8.915557e-01 	      2.589234e-01   8.442147e-01
 1.650000e+02     2.394199e-01   8.970581e-01   2.606861e-01   9.236384e-01 	      2.476289e-01   9.067845e-01
 1.950000e+02     2.221061e-01   8.678943e-01   2.426340e-01   8.878115e-01 	      2.274401e-01   8.520512e-01
 2.250000e+02     2.107988e-01   8.899542e-01   2.310569e-01   9.083592e-01 	      2.045533e-01   8.943931e-01
 2.550000e+02     1.975804e-01   8.916033e-01   2.181338e-01   9.243217e-01 	      1.910711e-01   9.093655e-01
 2.850000e+02     1.873698e-01   9.007037e-01   2.094143e-01   9.339328e-01 	      1.878438e-01   8.501912e-01
 3.200000e+02     1.905969e-01   9.260619e-01   2.150517e-01   9.698704e-01 	      1.843579e-01   9.490830e-01
 3.600000e+02     1.930121e-01   9.224827e-01   2.058315e-01   9.243900e-01 	      1.805468e-01   8.727121e-01
 4.000000e+02     1.940170e-01   9.113887e-01   2.200092e-01   9.000831e-01 	      1.832096e-01   9.347195e-01
 4.400000e+02     1.912622e-01   8.822357e-01   2.165050e-01   8.924887e-01 	      1.872435e-01   9.126458e-01
 4.800000e+02     1.809128e-01   9.888428e-01   2.019603e-01   1.022583e+00 	      1.778405e-01   8.522161e-01
}\ductE

\addplot [blue,semithick] table [x={0},y expr=\thisrow{1}/\thisrow{2}]{\ductE};
\addplot [darkgreen,semithick] table [x={0},y expr=\thisrow{3}/\thisrow{4}]{\ductE};
\addplot [red,semithick] table [x={0},y expr=\thisrow{5}/\thisrow{6}]{\ductE};

\node at (350.0,0.8) {\fcolorbox{black}{white}{\small{$5 < y_{12} < 6$}}};
\end{groupplot}
\end{tikzpicture}
\end{center}
\caption{
Gap fraction $f$ calculated with the $N_{\mi\pi} = 0$, with $N_{\mi\pi} = 2$, and with $\cV_{\mi\pi}$ exponentiated, ($N_{\mi\pi} = \infty$). We use maximum color suppression index $I_\textrm{max} = 4$ and $N_\Delta^\mathrm{thr} = 1$.
}
\label{fig:gap}
\end{figure}
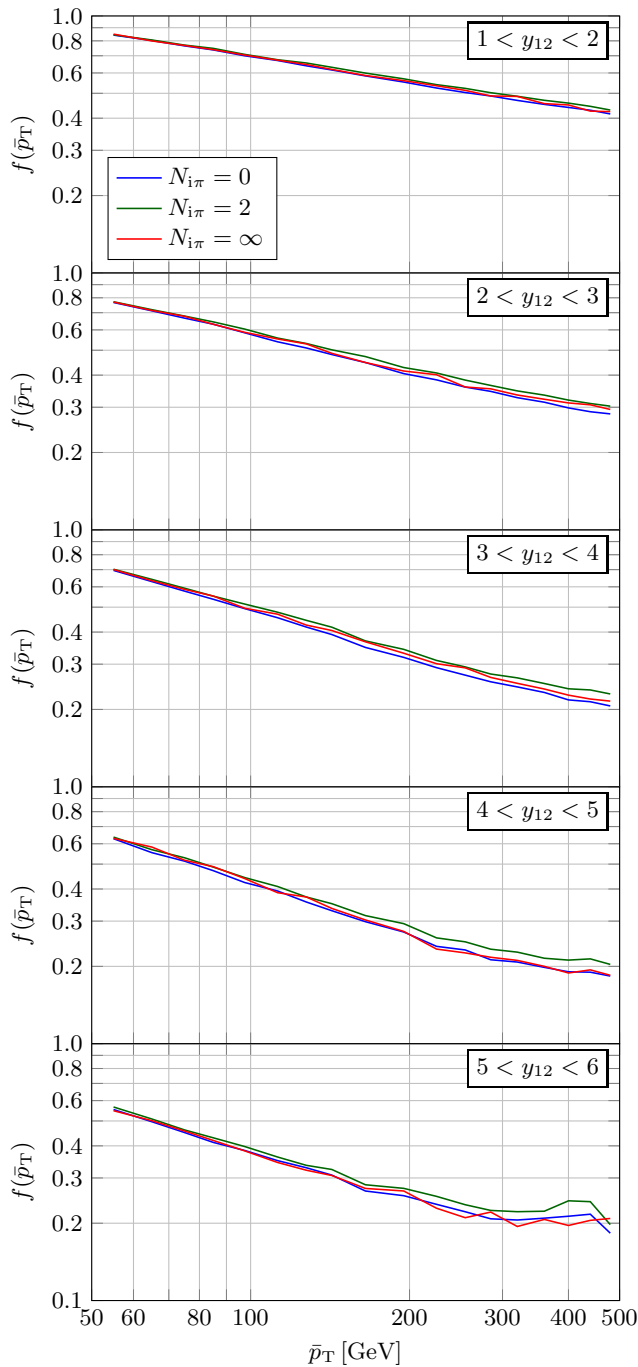

We choose five different bins for $y_{12}$ and examine $f(\bar p_\LT)$ as a function of $\bar p_\LT$. The results are shown in Fig.~\ref{fig:gap}. In each $y_{12}$ bin, we show three curves. The first, in blue, is obtained with $N_{\mi\pi} = 0$. Then, in green, we show results obtained with $N_{\mi\pi} = 2$. Finally, in red, we show the result with $\cV_{\mi\pi}$ in exponentiated form, $N_{\mi\pi} = \infty$.  The result with up to two powers of $\cV_{\mi\pi}$ ($N_{\mi\pi} = 2$) is generally somewhat larger than that with exponentiated $\cV_{\mi\pi}$, but the difference is typically not larger than 0.02.

It is remarkable that $f(\bar p_\LT, y_{12})$ with $\cV_{\mi\pi}$ in exponentiated form is always quite close to its value with $N_{\mi\pi} = 0$. This is consistent with what we found in Ref.~\cite{NSGapColor} and in Fig.~\ref{fig:gap}: allowing two powers of $\cV_{\mi\pi}$ perturbatively makes a rather small difference in the gap fraction. This does not mean that color beyond the LC+ approximation is unimportant for the gap fraction. We found in Ref.~\cite{NSGapColor} that allowing up to two powers of $\Delta \cH$ and $\Delta \cV_\mathrm{Re}$ can make a substantial difference when $\log(\bar p_\LT/p_\LT^\mathrm{cut})$ and $y_{12}$ are large.

\section{\label{sec:conclusions}Conclusions}

The accuracy of leading order parton shower programs could be enhanced by improving the treatment of QCD color \cite{PlatzerSjodahl, Seymour2018, PlatzerSjodahlThoren, Isaacson:2018zdi, Forshaw:2019ver}. We have worked to make our program, \textsc{Deductor}, more accurate with respect to color \cite{NScolor, NSColorEffects, NSColorTheory, NSGapColor}. In the organization of \textsc{Deductor}, there are three operators that have a color structure that cannot be incorporated into the LC+ approximation that \textsc{Deductor} uses: $\Delta \cH$, $\Delta \cV_\mathrm{Re}$, and $\cV_{\mi\pi}$.

We found in Ref.~\cite{NSColorTheory} that it is possible to include the operators $\Delta \cH$, $\Delta \cV_\mathrm{Re}$, and $\cV_{\mi\pi}$ perturbatively. That is, one can work with the result expanded in powers of these operators and retain the contributions up to some number of powers. There are practical limitations on the number of powers that one can retain, but two powers turns out to be quite practical.

The operator $\cV_{\mi\pi}$, which corresponds to the imaginary part of one-loop virtual graphs, is of special interest because of the role that it plays in analytical summation of large logarithms and because it contains a factor of $4\pi$, which makes it a numerically large operator. One wonders if one could include $\cV_{\mi\pi}$ to all orders. That is, one wonders if one could use $\cV_{\mi\pi}$ as part of the exponentiated Sudakov factor that occurs between real emissions.

In this paper, we have found that one can include $\cV_{\mi\pi}$ in the Sudakov exponent and we show how to do that. One has to calculate numerically the exponential of matrices, but the dimensionality of the matrices is at most $14\times 14$.

In principle, one could include $\cV_{\mi\pi}$ in the Sudakov exponent while including powers of $\Delta \cH$ and $\Delta \cV_\mathrm{Re}$ perturbatively. This would, however, require substantial effort to realize in computer code, so we have not attempted this further step.

We have tested this summation numerically, using the rapidity gap survival probability, which we have found is generally sensitive to color \cite{NSGapColor}. For the numerical test, we compare the results from the perturbative approach to the result with $\cV_{\mi\pi}$ included in the Sudakov exponent. The result of the two calculations agree that the effect of $\cV_{\mi\pi}$ is small, near the limit of what we can measure numerically.

We have also examined the effect of $\cV_{\mi\pi}$, both in exponentiated form and perturbatively, for a wide range of the variables $\bar p_\LT$ and $y_{12}$ that appear in the rapidity gap problem. We find that the effect of $\cV_{\mi\pi}$ is always quite small. 

It is perhaps not a surprise that the net $\cV_{\mi\pi}$ contribution is quite small. We note that if we integrate Eq.~(\ref{eq:Vipi}) for $\cV_{\mi\pi}$ over a range of scales from $p_\LT^\mathrm{cut} = 20 \GeV$ to $\bar p_\LT = 300 \GeV$ and leave out the color factors, we have a net phase $\phi \sim 2 \as \log((300/20)^2)$. With $\as \sim 0.1$, this is $\phi \sim 1$. Thus $\cV_{\mi\pi}$ is not a small operator in the context of the rapidity gap problem. However the contributions from $\cV_{\mi\pi}$ have opposite signs when acting on ket color states compared to the sign when acting on bra color states. We note that for real numbers $a$ and $\phi$, $\exp(a + \mi \phi) \exp(a - \mi \phi)$ is just $\exp(2a)$, with no contribution from the phase $\phi$. Thus the effect of $\cV_{\mi\pi}$ has to arise because the relevant color matrices don't commute. In our numerical tests of the perturbative approach to color in Ref.~\cite{NSColorTheory}, we found that the effects of $\cV_{\mi\pi}$ separately on the bra and ket states were of order 1, but that these effects cancelled, as they should, when we examined the unitarity property of shower evolution. Only when we have a color sensitive observable can the contributions fail to cancel. Apparently, the gap fraction is not sufficiently color sensitive to prevent a high degree of cancellation.

Because the effect of $\cV_{\mi\pi}$ is small for the gap fraction observable, we believe that it suffices to examine observables that might be sensitive to color by treating $\cV_{\mi\pi}$ perturbatively, along with $\Delta \cH$ and $\Delta \cV_\mathrm{Re}$. If one finds an observable in which $\cV_{\mi\pi}$ has a numerically substantial effect, then we could consider exponentiating this effect in \textsc{Deductor}. Of course, other workers may want to use the method of this paper to include $\cV_{\mi\pi}$ as part of the Sudakov exponent in other parton shower programs.

\acknowledgments{ 
This work was supported in part by the United States Department of Energy under grant DE-SC0011640.
}

\vfil


\begin{thebibliography}{99}


 \bibitem{NScolor}
  Z.~Nagy and D.~E.~Soper,
  {\em Parton shower evolution with subleading color},
  \href{http://dx.doi.org/10.1007/JHEP06(2012)044}
  {JHEP {\bf 1206}, 044 (2012)}
  [\href{http://inspirehep.net/search?p=find+doi+10.1007/JHEP06(2012)044}
  {\textsc{inSPIRE}}].

\bibitem{PlatzerSjodahl} 
  S.~Pl\"atzer and M.~Sj\"odahl,
  {\em Subleading $N_c$ improved Parton Showers},
  \href{http://dx.doi.org/10.1007/JHEP07(2012)042}
  {JHEP {\bf 1207}, 042 (2012)}
  [\href{http://inspirehep.net/search?p=find+doi+10.1007/JHEP07(2012)042}
  {\textsc{inSPIRE}}].
  
\bibitem{NSColorEffects}
  Z.~Nagy and D.~E.~Soper,
  {\em Effects of subleading color in a parton shower},
  \href{http://dx.doi.org/10.1007/JHEP07(2015)119}
  {JHEP {\bf 1507}, 119 (2015)}
  [\href{http://inspirehep.net/search?p=find+doi+10.1007/JHEP07(2015)119}
  {\textsc{inSPIRE}}].

\bibitem{Seymour2018}
  R.~\'Angeles Mart\'inez, M.~De Angelis, J.~R.~Forshaw, S.~Pl\"atzer 
  and M.~H.~Seymour,
  {\em Soft gluon evolution and non-global logarithms},
  \href{http://dx.doi.org/10.1007/JHEP05(2018)044}
  {JHEP {\bf 1805}, 044 (2018)}
  [\href{http://inspirehep.net/search?p=find+doi+10.1007/JHEP05(2018)044}
  {\textsc{inSPIRE}}].
  
\bibitem{PlatzerSjodahlThoren} 
  S.~Pl\"atzer, M.~Sj\"odahl and J.~Thor\'en,
  {\em Color matrix element corrections for parton showers},
  \href{http://dx.doi.org/10.1007/JHEP11(2018)009}
  {JHEP {\bf 1811}, 009 (2018)}
  [\href{http://inspirehep.net/search?p=find+doi+10.1007/JHEP11(2018)009}
  {\textsc{inSPIRE}}].

\bibitem{Isaacson:2018zdi} 
  J.~Isaacson and S.~Prestel,
  {\em Stochastically sampling color configurations},
  \href{http://dx.doi.org/10.1103/PhysRevD.99.014021}
  {Phys.\ Rev.\ D {\bf 99}, 014021 (2019)}
  [\href{http://inspirehep.net/search?p=find+doi+10.1103/PhysRevD.99.014021}
  {\textsc{inSPIRE}}].

\bibitem{Forshaw:2019ver} 
  J.~R.~Forshaw, J.~Holguin and S.~Pl\"atzer,
  {\em Parton branching at amplitude level},
  arXiv:1905.08686 [hep-ph]
  [\href{http://inspirehep.net/record/1735804}
  {\textsc{inSPIRE}}].
  
\bibitem{NSColorTheory} 
  Z.~Nagy and D.~E.~Soper,
  {\em Parton showers with more exact color evolution},
  \href{http://dx.doi.org/10.1103/PhysRevD.99.054009}
  {Phys.\ Rev.\ D {\bf 99}, 054009 (2019)}
  [\href{http://inspirehep.net/search?p=find+doi+10.1103/PhysRevD.99.054009}
  {\textsc{inSPIRE}}].

\bibitem{NSGapColor}
  Z.~Nagy and D.~E.~Soper,
  {\em Effect of color on rapidity gap survival},
  arXiv:1905.07176 [hep-ph]
  [\href{http://inspirehep.net/record/1735363}
  {\textsc{inSPIRE}}].


\bibitem{AtlasGap}
  G.~Aad {\it et al.} [ATLAS Collaboration],
  {\em Measurement of dijet production with a veto on additional central 
  jet activity in $pp$ collisions at $\sqrt{s}=7$ TeV using the ATLAS detector},
  \href{http://dx.doi.org/10.1007/JHEP09(2011)053}
  {JHEP {\bf 1109}, 053 (2011)}
  [\href{http://inspirehep.net/search?p=find+doi+10.1007/JHEP09(2011)053}
  {\textsc{inSPIRE}}].

\bibitem{HEJgap} 
  S.~Alioli, J.~R.~Andersen, C.~Oleari, E.~Re and J.~M.~Smillie,
  {\em Probing higher-order corrections in dijet production at the LHC},
  \href{http://dx.doi.org/10.1103/PhysRevD.85.114034}
  {Phys.\ Rev.\ D {\bf 85}, 114034 (2012)}
  [\href{http://inspirehep.net/search?p=find+doi+10.1103/PhysRevD.85.114034}
  {\textsc{inSPIRE}}]

\bibitem{HocheSchonherr} 
  S.~Hoeche and M.~Schonherr,
  {\em Uncertainties in next-to-leading order plus parton shower matched 
  simulations of inclusive jet and dijet production},
  \href{http://dx.doi.org/10.1103/PhysRevD.86.094042}
  {Phys.\ Rev.\ D {\bf 86}, 094042 (2012)}
  [\href{http://inspirehep.net/search?p=find+doi+10.1103/PhysRevD.86.094042}
  {\textsc{inSPIRE}}]


\bibitem{Manchester2009}
  J.~Forshaw, J.~Keates and S.~Marzani,
  {\em Jet vetoing at the LHC},
  \href{http://dx.doi.org/10.1088/1126-6708/2009/07/023}
  {JHEP {\bf 0907}, 023 (2009)}
  [\href{http://inspirehep.net/search?p=find+doi+10.1088/1126-6708/2009/07/023}
  {\textsc{inSPIRE}}].

\bibitem{EarlyGap1}
  N.~Kidonakis, G.~Oderda and G.~F.~Sterman,
  {\em Evolution of color exchange in QCD hard scattering},
  \href{http://dx.doi.org/10.1016/S0550-3213(98)00441-6}
  {Nucl.\ Phys.\ B {\bf 531}, 365 (1998)}
  [\href{http://inspirehep.net/search?p=find+doi+10.1016/S0550-3213(98)00441-6}
  {\textsc{inSPIRE}}].
  
\bibitem{EarlyGap2}
  G.~Oderda and G.~F.~Sterman,
  {\em Energy and color flow in dijet rapidity gaps},
  \href{http://dx.doi.org/10.1103/PhysRevLett.81.3591}
  {Phys.\ Rev.\ Lett.\  {\bf 81}, 3591 (1998)}
  [\href{http://inspirehep.net/search?p=find+doi+10.1103/PhysRevLett.81.3591}
  {\textsc{inSPIRE}}].

\bibitem{Manchester2005}
  J.~R.~Forshaw, A.~Kyrieleis and M.~H.~Seymour,
  {\em Gaps between jets in the high energy limit},
  \href{http://dx.doi.org/10.1088/1126-6708/2005/06/034}
  {JHEP {\bf 0506}, 034 (2005)}
  [\href{http://inspirehep.net/search?p=find+doi+10.1088/1126-6708/2005/06/034}
  {\textsc{inSPIRE}}].

\bibitem{NonGlobal1}
  M.~Dasgupta and G.~P.~Salam,
  {\em Resummation of nonglobal QCD observables},
  \href{http://dx.doi.org/10.1016/S0370-2693(01)00725-0}
  {Phys.\ Lett.\ B {\bf 512}, 323 (2001)}
  [\href{http://inspirehep.net/search?p=find+doi+10.1016/S0370-2693(01)00725-0}
  {\textsc{inSPIRE}}]

\bibitem{NonGlobal2}
  C.~F.~Berger, T.~Kucs and G.~F.~Sterman,
  {\em Energy flow in interjet radiation},
  \href{http://dx.doi.org/10.1103/PhysRevD.65.094031}
  {Phys.\ Rev.\ D {\bf 65}, 094031 (2002)}
  [\href{http://inspirehep.net/search?p=find+doi+10.1103/PhysRevD.65.094031}
  {\textsc{inSPIRE}}].

\bibitem{NonGlobal3}
  M.~Dasgupta and G.~P.~Salam,
  {\em Accounting for coherence in interjet E(t) flow: A Case study},
  \href{http://dx.doi.org/10.1088/1126-6708/2002/03/017}
  {JHEP {\bf 0203}, 017 (2002)}
  [\href{http://inspirehep.net/search?p=find+doi+10.1088/1126-6708/2002/03/017}
  {\textsc{inSPIRE}}].
  
\bibitem{NonGlobal4}
  R.~B.~Appleby and M.~H.~Seymour,
  {\em Nonglobal logarithms in interjet energy flow with kt clustering requirement},
  \href{http://dx.doi.org/10.1088/1126-6708/2002/12/063}
  {JHEP {\bf 0212}, 063 (2002)}
  [\href{http://inspirehep.net/search?p=find+doi+10.1088/1126-6708/2002/12/063}
  {\textsc{inSPIRE}}].
 
\bibitem{SuperLeading1}
  J.~R.~Forshaw, A.~Kyrieleis and M.~H.~Seymour,
  {\em Super-leading logarithms in non-global observables in QCD?},
  \href{http://dx.doi.org/10.1088/1126-6708/2006/08/059}
  {JHEP {\bf 0608}, 059 (2006)}
  [\href{http://inspirehep.net/search?p=find+doi+10.1088/1126-6708/2006/08/059}
  {\textsc{inSPIRE}}].

\bibitem{SuperLeading2}
  J.~R.~Forshaw, A.~Kyrieleis and M.~H.~Seymour,
  {\em Super-leading logarithms in non-global observables in QCD: 
  Colour basis independent calculation},
  \href{http://dx.doi.org/10.1088/1126-6708/2008/09/128}
  {JHEP {\bf 0809}, 128 (2008)}
  [\href{http://inspirehep.net/search?p=find+doi+10.1088/1126-6708/2008/09/128}
  {\textsc{inSPIRE}}].

\bibitem{Manchester2011}
  R.~M.~Duran Delgado, J.~R.~Forshaw, S.~Marzani and M.~H.~Seymour,
  {\em The dijet cross section with a jet veto}
  \href{http://dx.doi.org/10.1007/JHEP08(2011)157}
  {JHEP {\bf 1108}, 157 (2011)}
  [\href{http://inspirehep.net/search?p=find+doi+10.1007/JHEP08(2011)157}
  {\textsc{inSPIRE}}].


\bibitem{NSI}
  Z.~Nagy and D.~E.~Soper,
  {\em Parton showers with quantum interference},
  \href{http://dx.doi.org/10.1088/1126-6708/2007/09/114}
  {JHEP {\bf 0709}, 114 (2007)}
  [\href{http://inspirehep.net/search?p=find+doi+10.1088/1126-6708/2007/09/114}
  {\textsc{inSPIRE}}].

\bibitem{NSAllOrder} 
  Z.~Nagy and D.~E.~Soper,
  {\em What is a parton shower?},
  \href{http://dx.doi.org/10.1103/PhysRevD.98.014034}
  {Phys.\ Rev.\ D {\bf 98}, 014034 (2018)}
  [\href{http://inspirehep.net/search?p=find+doi+10.1103/PhysRevD.98.014034}
  {\textsc{inSPIRE}}].
  
\bibitem{Magnus:1954zz} 
  W.~Magnus,
  {\em On the exponential solution of differential equations 
  for a linear operator},
  \href{http://dx.doi.org/10.1002/cpa.3160070404}
  {Commun.\ Pure Appl.\ Math.\  {\bf 7}, 649 (1954)}
  [\href{http://inspirehep.net/search?p=find+doi+10.1002/cpa.3160070404}
  {\textsc{inSPIRE}}].

\bibitem{antikt}
  M.~Cacciari, G.~P.~Salam and G.~Soyez,
  {\em The Anti-k(t) jet clustering algorithm},
  \href{http://dx.doi.org/10.1088/1126-6708/2008/04/063}
  {JHEP {\bf 0804}, 063 (2008)}
  [\href{http://inspirehep.net/search?p=find+doi+10.1088/1126-6708/2008/04/063}
  {\textsc{inSPIRE}}].

\bibitem{NSThresholdII}
  Z.~Nagy and D.~E.~Soper,
  {\em Jets and threshold summation in Deductor},
  \href{http://dx.doi.org/10.1103/PhysRevD.98.014035}
  {Phys.\ Rev.\ D {\bf 98}, 014035 (2018)}
  [\href{http://inspirehep.net/search?p=find+doi+10.1103/PhysRevD.98.014035}
  {\textsc{inSPIRE}}].
  


\end{thebibliography}
\end{document}